\documentclass[12pt,a4paper]{article}
\usepackage[utf8]{inputenc}
\usepackage{epsf}
\usepackage{latexsym,amssymb,euscript}
\usepackage[dvips]{graphicx}
\usepackage[numbers,sort&compress]{natbib}
\usepackage{amsmath}
\usepackage{slashed}
\usepackage{booktabs}
\usepackage{hyperref}
\usepackage{braket}
\usepackage{chngcntr}
\usepackage{bbold}
\usepackage{graphics}
\usepackage{graphicx}
\usepackage{caption}
\usepackage{subcaption}
\usepackage{pdfpages}
\usepackage[titletoc]{appendix}
\graphicspath{{./figures/}}
\hypersetup{
 linktocpage = true,
 linkcolor = blue,
 urlcolor = blue,
 colorlinks = true,
 linkcolor = teal,
 anchorcolor = blue,
 citecolor = teal,
 urlcolor = teal,
 pdfstartview = {XYZ null null 1.25} 
           }
\usepackage[left=2cm, right=2cm]{geometry}
\usepackage{pstricks}
\usepackage{color}
\usepackage{float}
\usepackage{academicons}
\definecolor{orcidlogocol}{HTML}{A6CE39}

\newcommand{\mQp}{M_{Q \perp}}
\newcommand{\DY}{\Delta Y}
\newcommand{\drv}{{\rm d}}
\newcommand{\tild}[1]{~}

\newcommand{\tcite}[1]{~\cite{#1}}
\newcommand{\tref}[1]{~\ref{#1}}

\newcommand{\orcidADB}{\href{https://orcid.org/0000-0002-6114-7044}{\includegraphics[scale=0.1]{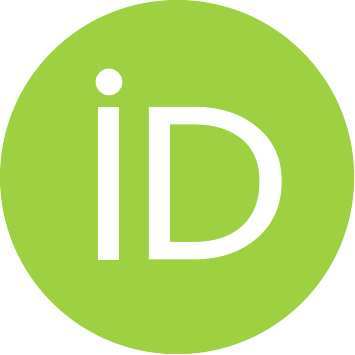}}}

\newcommand{\orcidFGC}{\href{https://orcid.org/0000-0003-3299-2203}{\includegraphics[scale=0.1]{figures/logo-orcid.pdf}}}

\newcommand{\orcidMF}{\href{https://orcid.org/0000-0002-2408-2210}{\includegraphics[scale=0.1]{figures/logo-orcid.pdf}}}

\newcommand{\orcidDYI}{\href{https://orcid.org/0000-0001-5701-4364}{\includegraphics[scale=0.1]{figures/logo-orcid.pdf}}}

\newcommand{\orcidAP}{\href{https://orcid.org/0000-0001-8984-3036}{\includegraphics[scale=0.1]{figures/logo-orcid.pdf}}}

\begin{document}

\begin{titlepage}

\begin{center}
  {\Large \bf Inclusive production of a heavy-light dijet system \\ in hybrid high-energy and collinear factorization}
\end{center}

\vskip 0.3cm

\centerline{
A.D.~Bolognino$^{1,2,*}$ \orcidADB,
F.G.~Celiberto$^{3,4,5,\dagger}$ \orcidFGC,
M.~Fucilla$^{1,2\ddagger}$ \orcidMF, 
D.Yu.~Ivanov$^{6,7,\S}$ \orcidDYI, 
and A.~Papa$^{1,2,\P}$ \orcidAP}

\vskip .4cm

\centerline{${}^1$ {\sl Dipartimento di Fisica, Universit\`a della Calabria,}}
\centerline{\sl I-87036 Arcavacata di Rende, Cosenza, Italy}
\vskip .18cm
\centerline{${}^2$ {\sl Istituto Nazionale di Fisica Nucleare, Gruppo collegato di Cosenza,}}
\centerline{\sl I-87036 Arcavacata di Rende, Cosenza, Italy}
\vskip .18cm
\centerline{${}^3$ {\sl European Centre for Theoretical Studies in Nuclear Physics and Related Areas (ECT*),}}
\centerline{\sl I-38123 Villazzano, Trento, Italy}
\vskip .18cm
\centerline{${}^4$ {\sl Fondazione Bruno Kessler (FBK), 
I-38123 Povo, Trento, Italy} }
\vskip .20cm
\centerline{${}^5$ {\sl INFN-TIFPA Trento Institute of Fundamental Physics and Applications,}}
\centerline{\sl I-38123 Povo, Trento, Italy}
\vskip .18cm
\centerline{${}^6$ {\sl Sobolev Institute of Mathematics, 630090 Novosibirsk,
    Russia}}
\vskip .18cm
\centerline{${}^7$ {\sl Novosibirsk State University, 630090 Novosibirsk,
    Russia}}
\vskip 0.50cm

\begin{abstract}
\vspace{0.25cm}
\hrule \vspace{0.50cm}
We propose the study of the inclusive hadroproduction of a heavy-flavored jet in association with a light jet, as a probe channel of strong interactions at high energies. We build up a hybrid factorization that encodes genuine high-energy effects, provided by a partial next-to-leading BFKL resummation, inside the standard collinear structure of the cross section. We present a detailed analysis of different distributions, shaped on kinematic ranges typical of experimental analyses at the Large Hadron Collider, and differential in rapidity, azimuthal angle and transverse momentum. The fair stability that these distributions exhibit under higher-order corrections motivates our interest toward future studies. Here, the hybrid factorization could help to deepen our understanding of heavy-flavor physics in wider kinematic ranges, like the ones accessible at the Electron-Ion Collider.
\vspace{0.50cm} \hrule
\vspace{0.50cm}
{
 \setlength{\parindent}{0pt}
 \textsc{Keywords}: QCD phenomenology, high-energy resummation, heavy flavor, jets
}
\end{abstract}

\vfill
$^{*}${\it e-mail}:
\href{mailto:ad.bolognino@unical.it}{ad.bolognino@unical.it}

$^{\dagger}${\it e-mail}:
\href{mailto:fceliberto@ectstar.eu}{fceliberto@ectstar.eu}

$^{\ddagger}${\it e-mail}:
\href{mailto:michael.fucilla@unical.it}{michael.fucilla@unical.it}

$^{\S}${\it e-mail}:
\href{mailto:d-ivanov@math.nsc.ru}{d-ivanov@math.nsc.ru}

$^{\P}${\it e-mail}:
\href{alessandro.papa@fis.unical.it}{alessandro.papa@fis.unical.it}

\end{titlepage}

\section{Introductory remarks}
\label{introduction}

Heavy-flavored emissions in hadronic and lepto-hadronic collisions are commonly recognized as excellent probe channels of the dynamics of strong interactions. This resulted in remarkable interest over the last decades on both their formal and phenomenological aspects.

The wide range of applications of heavy flavor to collinear physics (see, \emph{e.g.}, Ref.\tcite{Frixione:1997ma} and references therein) makes the following selection of results very short and incomplete.
In Refs.\tcite{Mele:1990yq,Mele:1990cw} a next-to-leading order (NLO) formalism for the fragmentation of heavy quarks was provided.
Then, the effect of higher-order soft-gluon resummation on heavy-flavor in hadroproduction was gauged\tcite{Bonciani:1998vc}.
In Ref.\tcite{Forte:2010ta}, a general framework for the inclusion of heavy-quark mass contributions to DIS structure functions, based on the so-called FONLL scheme\tcite{Cacciari:1998it}, was presented.
In Ref.\tcite{Ridolfi:2019bch} a study on the effect of collinear logarithms (genuinely emerging in heavy-flavored emissions) on fragmentation functions (FFs) was performed.
Quite recently, the impact of charm-tagged cross sections on the strange collinear parton distribution function (PDF) was weighed\tcite{Faura:2020oom}.
In Ref.\tcite{Catani:2020kkl}, QCD radiative corrections to the production of bottom-quark pairs in hadronic collisions were extended up to next-to-NLO.

Jet emissions in regimes dominated by \emph{transverse-momentum-dependent} (TMD) dynamics gave us a faultless chance to study single transverse-spin asymmetries\tcite{Bacchetta:2005rm,Bacchetta:2007sz,Qiu:2007ey,Vogelsang:2007jk}. Then, the formal description of dijet systems in hadroproduction channels made us understand that both the na\"ive and the generalized TMD factorization are violated\tcite{Collins:2007nk,Rogers:2010dm}. Photon-plus-jet final states were proposed\tcite{Buffing:2018ggv,Rogers:2013zha} as a suitable channel to gauge the size of these breaking effects. More recently, TMD factorization formulae for light-dijet\tcite{delCastillo:2020omr} and heavy-dijet\tcite{Kang:2020xgk} emissions in electron-proton scatterings were found in a soft-collinear effective scheme (SCET). A SCET approach was used to investigate the jet substructure and, more in particular, to define fragmenting jet functions, needed to describe the production of charmonia within jets\tcite{Procura:2009vm,Baumgart:2014upa,Bain:2017wvk}.

At small-$x$, a factorization formula was established\tcite{Catani:1990xk,Catani:1990eg,Collins:1991ty} for heavy-flavor production, in whose low-virtuality limit the standard collinear factorization is recovered.
In the saturation regime, where the size of nonlinear effects due to gluon recombination become significant, heavy-flavored states in hadronic collisions gave us access to gluon Wigner distributions (see, \emph{e.g.}, Ref.\tcite{Boussarie:2018zwg}).
On the phenomenological side, the impact of small-$x$ LHCb data sensitive to heavy-flavor production on PDFs was recently investigated in the fixed-flavor number scheme with NLO accuracy\tcite{Zenaiev:2015rfa}.

\begin{figure}[t]
\centering
\includegraphics[width=0.40\textwidth]{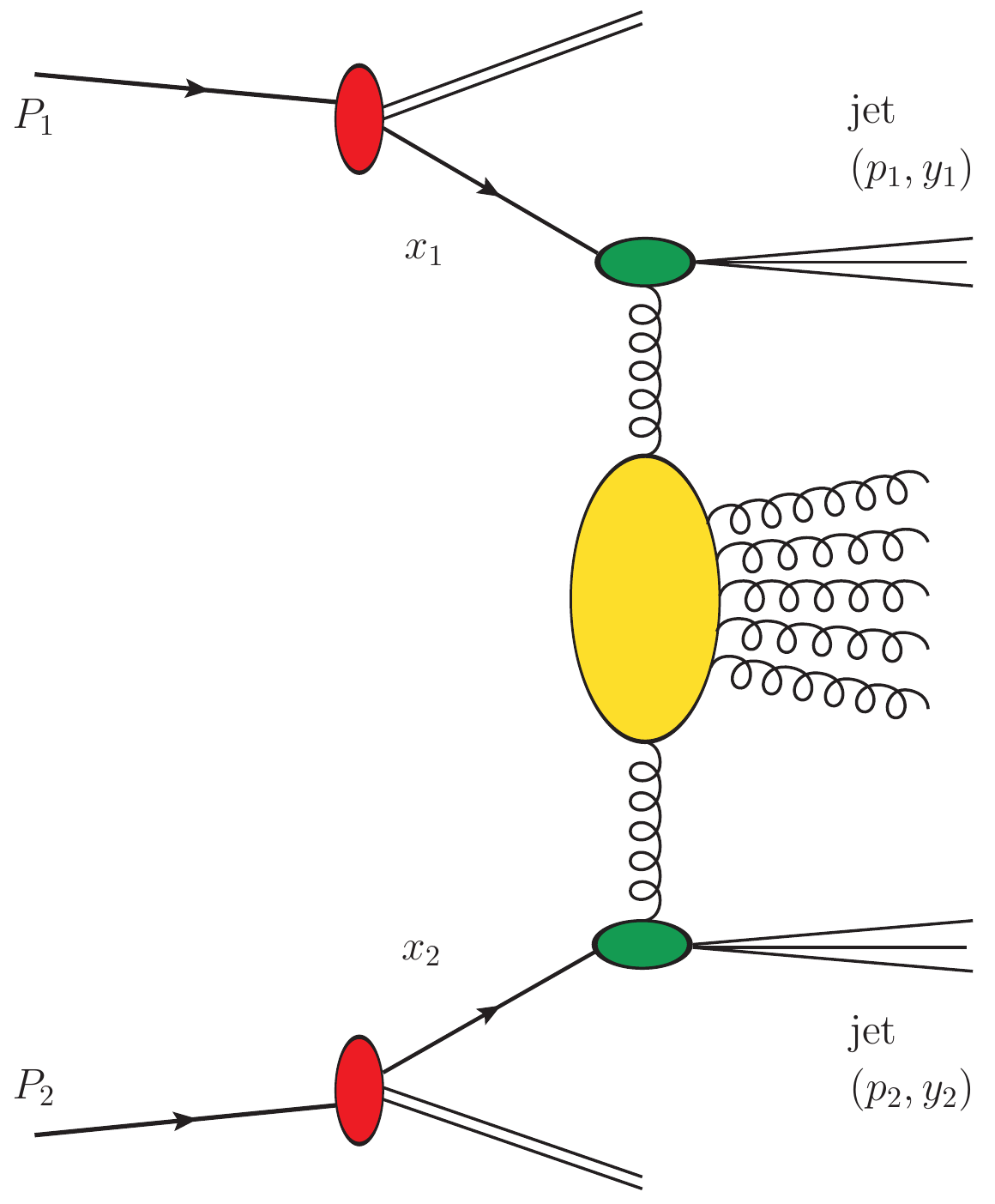}
\hspace{1.40cm}
\includegraphics[width=0.49\textwidth]{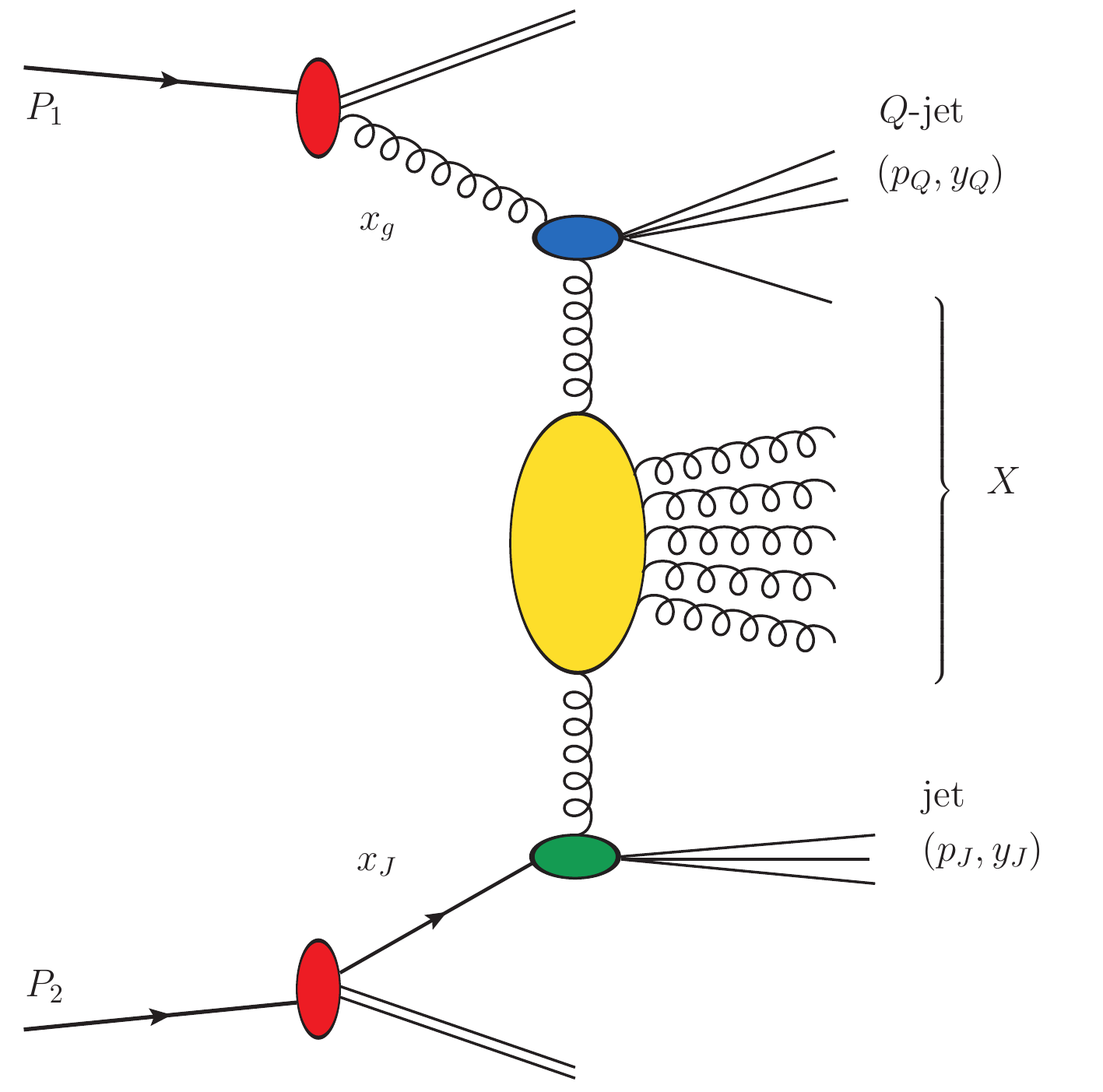}
\\ \vspace{0.25cm}
a) Mueller--Navelet jets \hspace{4.00cm}
b) Heavy-light dijet system
\caption{Pictorial representation of a) the Mueller--Navelet jet production and of b) the inclusive heavy-light dijet hadroproduction in hybrid high-energy/collinear factorization.}
\label{fig:process_2p}
\end{figure}

\begin{figure}[t]
\centering
\includegraphics[width=0.45\textwidth]{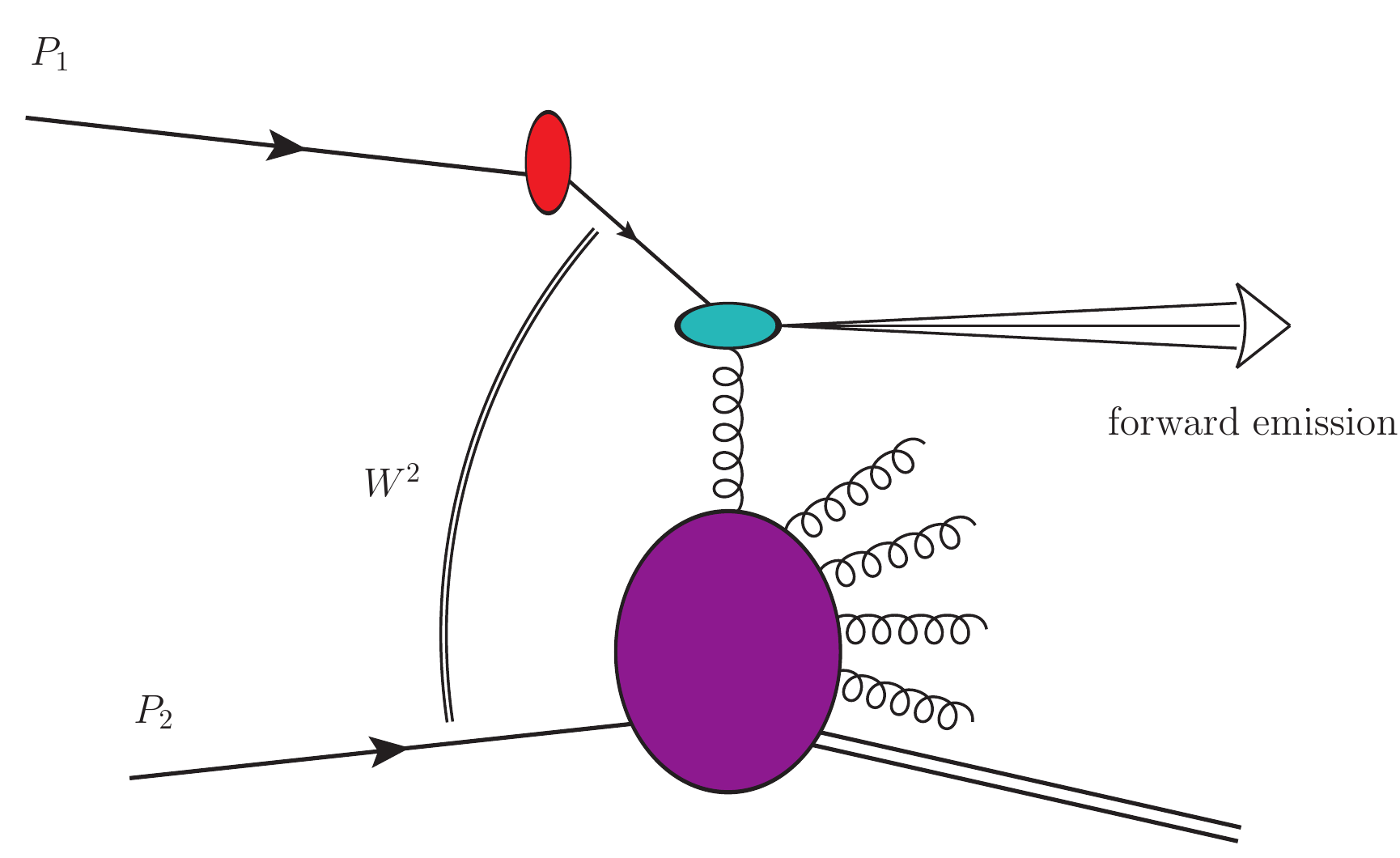}
\hspace{1.40cm}
\includegraphics[width=0.45\textwidth]{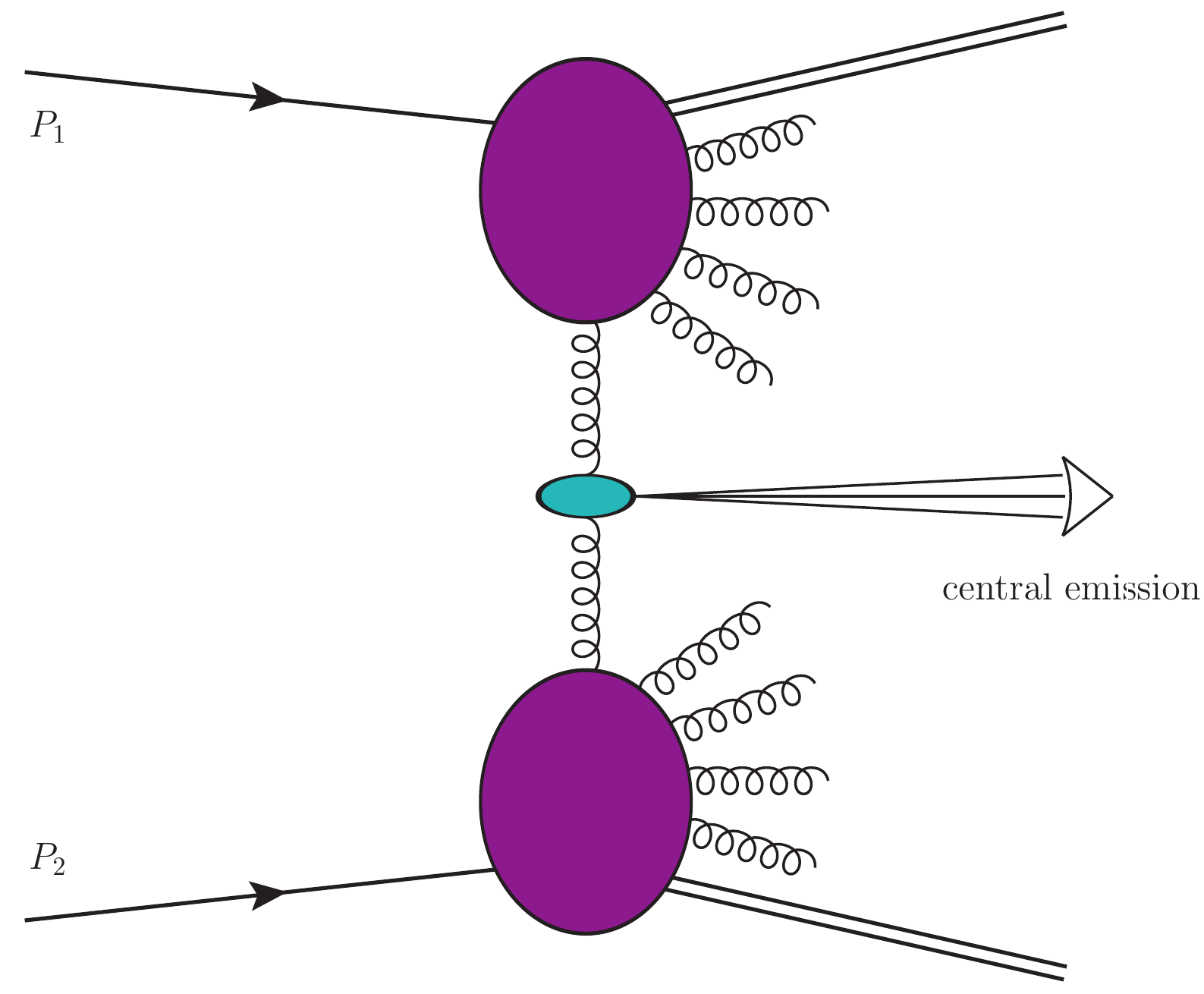}
\\ \vspace{0.25cm}
a) Single forward emission \hspace{4.00cm}
b) Single central emission
\caption{Pictorial representation of a) an inclusive single forward emission in hybrid high-energy/collinear factorization and of b) an inclusive single central emission in pure high-energy factorization. The red blob in panel a) refers to collinear PDFs, while sea green ones in both panels denote the hard part of the impact factor describing the emission of a generic object in a) forward or b) central regions of rapidity. The unintegrated gluon distributions, depicted in violet, encode nonperturbative information about the gluon content in the proton at high energies (small-$x$) and are connected to the impact factor via Reggeon lines. Gluon-induced emissions from the collinear region in panel a), not shown here, are encoded in the sea green blob.}
\label{fig:process_1p}
\end{figure}

\begin{figure}[t]
\centering
\includegraphics[width=0.65\textwidth]{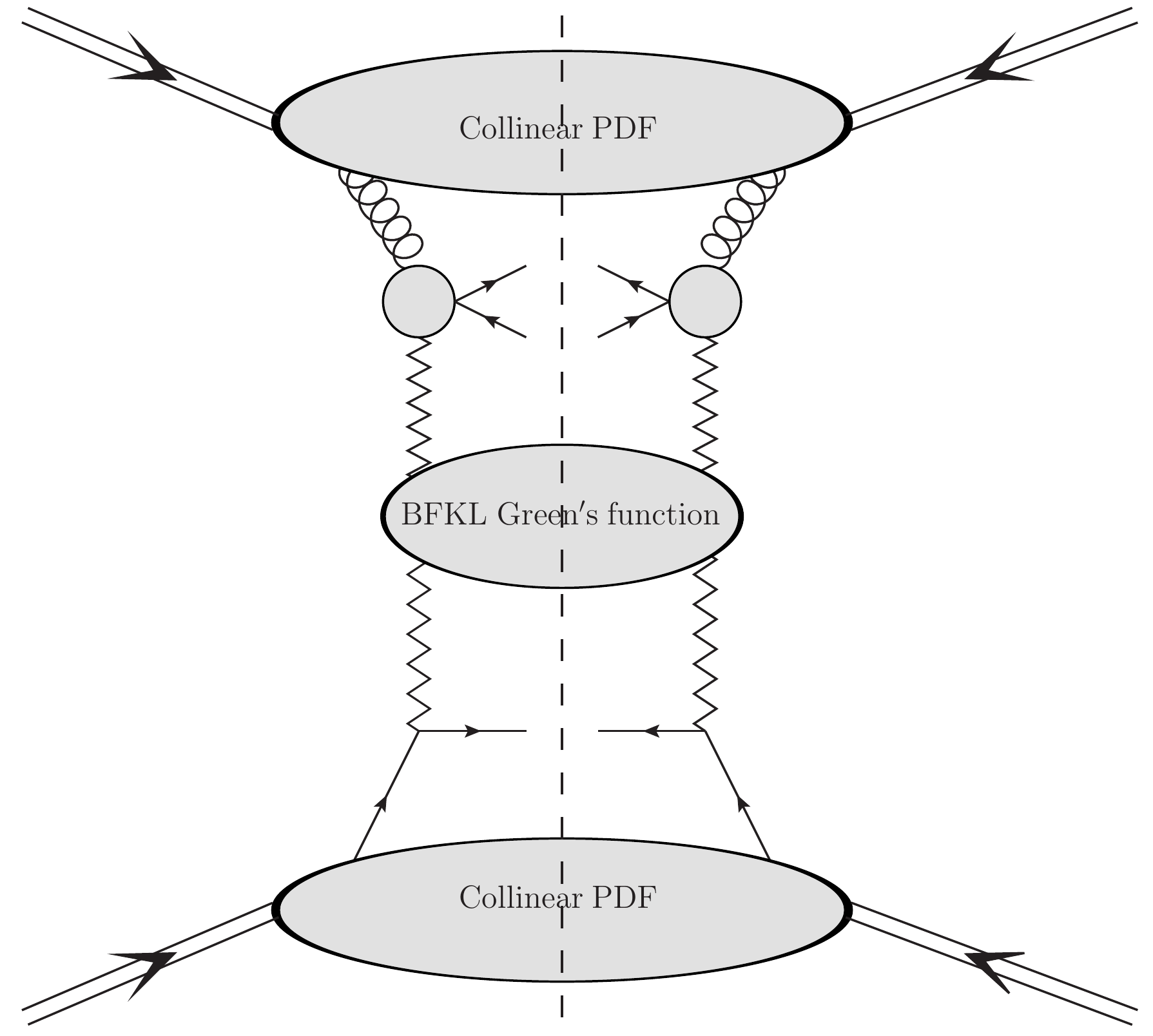}
\caption{Hybrid factorization for the heavy-light dijet production with a light-quark jet produced in the lower fragmentation region (gluon jet not shown here). The high-energy resummed partonic cross section is convoluted with collinear PDFs.}
\label{fig:process_factorization}
\end{figure}

\begin{figure}[t]
\centering
\includegraphics[width=0.25\textwidth]{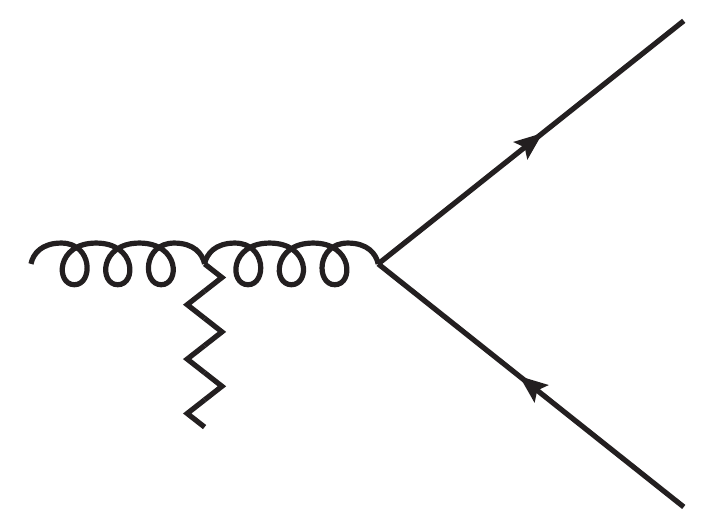}
\hspace{1.40cm}
\includegraphics[width=0.25\textwidth]{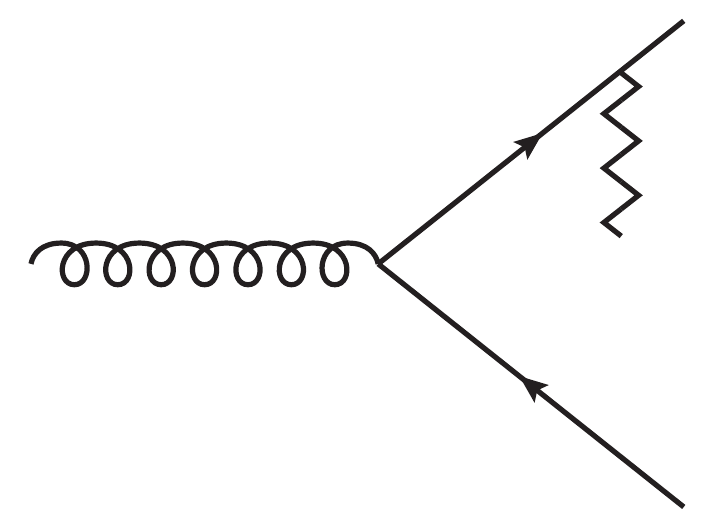}
\hspace{1.40cm}
\includegraphics[width=0.25\textwidth]{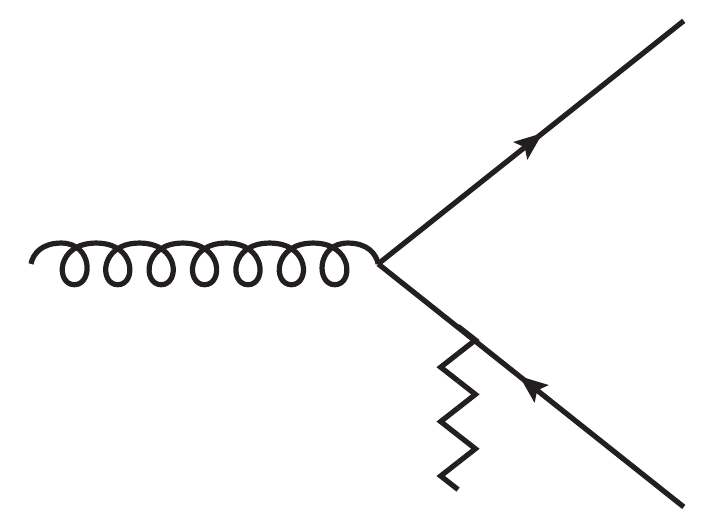}
\caption{Feynman diagrams relevant for the calculation of the LO impact factor describing the hadroproduction of a heavy-flavored jet. Zigzag lines stand for Reggeized gluons.}
\label{fig:hqp_IF}
\end{figure}

\begin{figure}[t]
\centering
\includegraphics[width=0.25\textwidth]{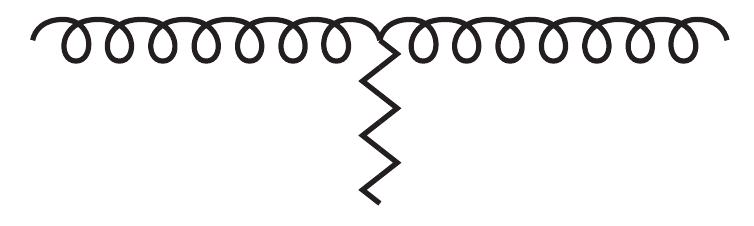}
\hspace{1.40cm}
\includegraphics[width=0.25\textwidth]{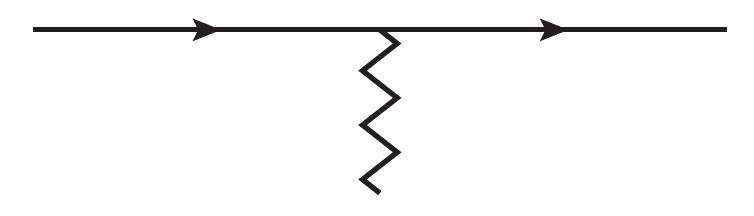}
\hspace{1.40cm}
\includegraphics[width=0.25\textwidth]{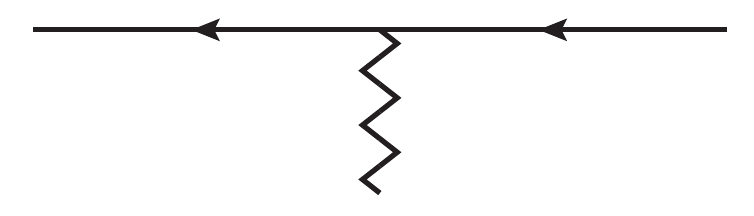}
\caption{Feynman diagrams relevant for the calculation of the LO impact factor describing the hadroproduction of a light jet. Zigzag lines stand for Reggeized gluons.}
\label{fig:jet_IF}
\end{figure}

In this work, our interest falls on processes involving inclusive heavy- and light-jet emissions at high energies.
Kinematic regimes under investigation generally lead to 
a \emph{semi-hard} scale ordering\tcite{Gribov:1984tu}, $\Lambda_{\rm QCD} \ll Q_i \ll \sqrt s$, where $\Lambda_{\rm QCD}$ is the QCD-coupling scale parameter, $Q_i$ indicates one or more, process-related hard scales, and $\sqrt s$ is the center-of-mass energy.
In this domain, a fixed-order treatment based on pure collinear factorization breaks down since large energy logarithms emerge in the perturbative series with a power increasing with the order.
More in particular, large final-state rapidities (or rapidity intervals), typical of diffractive semi-hard final states, strengthen the weight of terms proportional to $\ln (s)$.
The Balitsky--Fadin--Kuraev--Lipatov (BFKL) approach\tcite{Fadin:1975cb,Kuraev:1976ge,Kuraev:1977fs,Balitsky:1978ic} affords us an all-order resummation of these large energy logarithms both in the leading approximation (LLA), that includes all terms proportional to $\alpha_s^n \ln (s)^n$, and in the next-to-leading logarithmic approximation (NLA), that includes all terms proportional to $\alpha_s^n \ln (s)^{n-1}$.

Following the BFKL treatment, the analytic form of the imaginary part of the process scattering amplitude is expressed as a convolution of two \emph{impact factors}, portraying the transition from each colliding particle to the respective final-state object, and of a \emph{Green's function}. 
The latter is a process-independent object and it is ruled by an integral evolution equation, whose kernel is well-known at leading order (LO) as well as at NLO for forward scattering\tcite{Fadin:1998py,Ciafaloni:1998gs}, and also for any fixed, not growing with $s$, momentum transfer $t$ and for any possible two-gluon color state in the $t$-channel \tcite{Fadin:1998jv,Fadin:2000,Fadin:2005fiore,Fadin:2005RF}. 
\emph{Vice versa}, impact factors depend on the processes, thus representing the most challenging part of the calculation. So far, the list of impact factors calculated with NLO accuracy is quite short. Colliding-parton impact factors\tcite{Fadin:1999de,Fadin:1999df} represent the key ingredient to build the forward-jet\tcite{Bartels:2001ge,Bartels:2002yj,Caporale:2011cc,Caporale:2012ih,Ivanov:2012ms,Colferai:2015zfa} and the forward light-charged hadron\tcite{Ivanov:2012iv} impact factor. The ones describing the ($\gamma^* \to \gamma^*$) and the ($\gamma^* \to \mbox{light-vector meson}$) transition were respectively calculated in Refs.\tcite{Ivanov:2004pp} and\tcite{Bartels:2000gt,Bartels:2001mv,Bartels:2002uz,Bartels:2004bi,Fadin:2001ap,Balitsky:2012bs}.
Quite recently\tcite{Hentschinski:2020tbi}, a novel calculation of the forward-Higgs impact factor was done by using results on one-loop amplitudes\tcite{Nefedov:2017qzc,Nefedov:2018vyt,Nefedov:2019mrg} calculated via the high-energy effective theory developed by Lev N. Lipatov\tcite{Lipatov:1995pn,Lipatov:1996ts}.

Over the last two decades, a remarkable number for semi-hard reactions were proposed as sounding probes of the high-energy resummation (see Ref.\tcite{Celiberto:2017ius} for a recent review). An incomplete list of them includes: the dimeson exclusive electroproduction~\cite{Ivanov:2005gn,Ivanov:2006gt,Enberg:2005eq},
the ($\gamma*$-$\gamma*$) process~\cite{Ivanov:2014hpa},
the inclusive hadroproduction of a light dijet system emitted with large transverse momenta and a wide distance in rapidity (Mueller--Navelet jets~\cite{Mueller:1986ey}),
for which several phenomenological studies have been given~(see, \emph{e.g.},~Refs.~\cite{Colferai:2010wu,Caporale:2012ih,Ducloue:2013hia,Ducloue:2013bva,Caporale:2013uva,Caporale:2014gpa,Colferai:2015zfa,Caporale:2015uva,Ducloue:2015jba,Celiberto:2015yba,Celiberto:2015mpa,Celiberto:2016ygs,Celiberto:2016vva,Caporale:2018qnm,Chachamis:2015crx}), the inclusive multi-jet production~\cite{Caporale:2015vya,Caporale:2015int,Caporale:2016soq,Caporale:2016vxt,Caporale:2016xku,Celiberto:2016vhn,Caporale:2016djm,Caporale:2016lnh,Caporale:2016zkc}, the inclusive detection of a light-charged dihadron system~\cite{Celiberto:2016hae,Celiberto:2016zgb,Celiberto:2017ptm}, the heavy-quark pair hadro-~\cite{Bolognino:2019yls} and photoproduction~\cite{Celiberto:2017nyx,Bolognino:2019ouc}. Then, $J/\Psi$-plus-jet~\cite{Boussarie:2017oae},
hadron-plus-jet~\cite{Bolognino:2018oth,Bolognino:2019yqj,Bolognino:2019cac,Celiberto:2020rxb}, Drell--Yan-plus-jet~\cite{Golec-Biernat:2018kem,Deak:2018obv}, and Higgs-plus-jet~\cite{Celiberto:2020tmb} inclusive hadroproduction were proposed.

Another intriguing perspective is represented by the investigation of the hadronic structure in the high-energy limit via the BFKL approach.
Single forward emissions in exclusive as well as in inclusive channels give us direct access to the \emph{unintegrated gluon distribution} (UGD) in the proton, operationally defined in terms of a convolution between the BFKL Green’s function and a nonperturbative input, embodied in the proton impact.
First probes of the UGD were done in the context of deep-inelastic-scattering (DIS) structure functions\tcite{Hentschinski:2012kr,Hentschinski:2013id}. Then, the UGD has been tested through the single exclusive leptoproduction of $\rho$ and $\phi$ mesons\tcite{Anikin:2009bf,Anikin:2011sa,Besse:2013muy,Bolognino:2018rhb,Bolognino:2018mlw,Bolognino:2019bko,Bolognino:2019pba,Celiberto:2019slj} at HERA, the forward inclusive production of Drell--Yan lepton pairs\tcite{Motyka:2014lya,Brzeminski:2016lwh,Motyka:2016lta,Celiberto:2018muu} at LHCb, and the exclusive photoproduction of quarkonium states\tcite{Bautista:2016xnp,Garcia:2019tne,Hentschinski:2020yfm}. 
The connection between the UGD and the collinear gluon PDF has been studied in deep via a high-energy factorization framework set up in the early nineties\tcite{Catani:1990xk,Catani:1990eg,Collins:1991ty}, and in the Catani--Ciafaloni--Fiorani--Marchesini (CCFM) \emph{branching} scheme\tcite{Ciafaloni:1987ur,Catani:1989sg,Catani:1989yc,Marchesini:1994wr,Kwiecinski:2002bx}. Then, first determinations of small-$x$ improved PDFs \emph{\`a la} Altarelli--Ball--Forte (ABF)\tcite{Ball:1995vc,Ball:1997vf,Altarelli:2001ji,Altarelli:2003hk,Altarelli:2005ni,Altarelli:2008aj,White:2006yh} were recently proposed\tcite{Ball:2017otu,Abdolmaleki:2018jln,Bonvini:2019wxf}.
Conversely, the link between the BFKL UGD and TMD gluon distribution functions\tcite{Mulders:2000sh,Meissner:2007rx,Boer:2016xqr} is still more obscure. From the formal point of view, the TMD formalism is based on parton correlators (thus, on Wilson lines), while the BFKL approach ``speaks" the language of Reggeized gluons. From the phenomenological point of view, the TMD factorization is expected to work at low $p_T$, while the BFKL resummation requires large-$p_T$ emissions.
A first connection between the unpolarized and the linearly polarized gluon TMDs, $f^g_1$ and $h^{\perp g}_1$,
was established in the [low-$x$, large-$p_T$] limit\tcite{Dominguez:2011wm}.
More recently, a model calculation
of BFKL-improved unpolarized and polarized gluon TMDs was provided\tcite{Bacchetta:2020vty,Celiberto:2021zww}.

Concerning LHC phenomenology, one of the most popular semi-hard channels is the Mueller--Navelet jet production.
As already mentioned, final states here are characterized by two jets produced in proton-proton collisions, having large $p_T$ and being well separated in rapidity. Being the first totally hadroproduced reaction studied in the high-energy formalism, setting its formal description up required a substantial effort. On the one hand, the BFKL resummation was originally designed to study purely partonic (or lepto-hadronic) cross sections. On the other hand, kinematic configurations attainable at LHC detectors feature moderate values of parton $x$, this justifying a description in terms of collinear PDFs, but also large transverse $p_T$ exchanged in the $t$-channel, this calling for a $p_T$-factorized treatment, ensured by BFKL.
Therefore, a \emph{hybrid} high-energy/collinear factorization was built up, where high-energy resummed partonic cross sections are natively calculated in the BFKL approach, and then convoluted with collinear PDFs (see panel a) of Fig.\tref{fig:process_2p} for a schematic view).

The Mueller--Navelet channel can be considered as the ``father" of those processes falling in the class of inclusive forward/backward two-object emissions. There are other diffractive semi-hard final states that can be described by the hand of a hybrid factorization.
Inclusive single forward emissions in hadroprodocution reactions (panel a) of Fig.\tref{fig:process_1p}) lead to an asymmetric configuration where a parton always participates in the hard subprocess with a large $x$, while the other parton is a small-$x$ gluon. Here, a different kind of hybrid high-energy/collinear factorization is realized, so that the large-$x$ parton is portrayed by a collinear PDFs, whereas the small-$x$ gluon evolution is driven by a UGD. To get the hadronic cross section, the two densities are then convoluted with the impact factor depicting the forward-object emission. A representative process of this class is the forward Drell--Yan dilepton production\tcite{Motyka:2014lya,Celiberto:2018muu}.
Another formalism, close in spirit with our hybrid factorization for single forward emissions, has been employed first in the description of forward jets at the LHC\tcite{Deak:2009xt}, then in the study of $Z_0$-plus jet configurations\tcite{vanHameren:2015uia,Deak:2018obv} and, more recently, for the investigation of three-jet event topologies\tcite{VanHaevermaet:2020rro}.
In that case, \emph{partially off-shell} squared matrix elements correspond to our BFKL impact factors, whereas different models for the UGD have been used.

For completeness, we mention also gluon-induced inclusive single central emissions (panel b) of Fig.\tref{fig:process_1p}). At variance with the previous cases, here a pure high-energy factorization can be used, since both the initial-state gluons are extracted from the parent nucleons at small-$x$. Cross section is thus obtained as a convolution of two UGDs with a central-production impact factor, also known as \emph{off-shell coefficient function}. Distinctive reactions already studied in high-energy factorization are: the central-jet production, for which the NLO impact factor was calculated\tcite{Bartels:2006hg}, the single quarkonium emission\tcite{Baranov:2002cf,Kniehl:2006sk,Kniehl:2016sap,Cisek:2017gno,Cisek:2017ikn,Maciula:2018bex,Prokhorov:2020owf}, for which several analyses on the hadronization mechanism have been performed so far, and the central-Higgs radiation\tcite{Lipatov:2005at,Pasechnik:2006du,Abdulov:2017tis,Lipatov:2014mja}, where an exciting outlook would be the comparison between pure high-energy predictions and small-$x$ improved ABF results\tcite{Bonvini:2018ixe,Bonvini:2018iwt}.
The inclusive reactions presented in Figs.\tref{fig:process_2p} and\tref{fig:process_1p} are complemented by their exclusive counterparts, and by lepto-hadronic channels, not shown here.

Mueller--Navelet jet production is the only semi-hard process for which LHC data have been collected so far.
Here, a first theory-versus-experiment study on the azimuthal-angle correlation between the two light jets, complied by the CMS Collaboration at 7 TeV collision energy\tcite{Khachatryan:2016udy}, revealed that the kinematic domain under consideration stays in between the sectors described by the BFKL and the Dokshitzer--Gribov--Lipatov--Altarelli--Parisi (DGLAP)\tcite{Gribov:1972ri,Gribov:1972rt,Lipatov:1974qm,Altarelli:1977zs,Dokshitzer:1977sg} approach, while more evident high-energy signatures are expected to emerge at increasing energies. Then, recent analyses conducted at our Group\tcite{Celiberto:2015yba,Celiberto:2015mpa} highlighted how BFKL imprints can be efficaciously disengaged by the DGLAP background by considering asymmetric cuts for the $p_T$ of the two jets. This results was later confirmed also for the dihadron and hadron-jet production cases\tcite{Celiberto:2020wpk}.

It is generally known, however, that practical applications of the BFKL mechanism to physical reactions suffer from instabilities of the resummed series. It stems from the fact that higher-order (NLA) corrections are large, both in the kernel of the Green’s function and in the non-universal impact factors, and with opposite sign with respect to LLA terms. 
This translates in a raised sensitivity of the series on renormalization scale variation, which, in Mueller--Navelet case, is so strong to prevent any attempt to perform reliable analyses around ``natural" scales (which correspond, for light jets, to their $p_T$).
All these issues have to be handled by the use of some optimization procedure of the QCD perturbative series, and several studies done by adopting different optimization methods have been performed so far (see, \emph{e.g.}, Refs.\tcite{Vera:2006un,Vera:2007kn,Caporale:2013uva,Ducloue:2013hia,Ducloue:2013bva,Caporale:2014gpa,Caporale:2015uva}).
One of the most popular optimization scheme is the so-called Brodsky--Lepage--Mackenzie (BLM) method \tcite{Brodsky:1996sg,Brodsky:1997sd,Brodsky:1998kn,Brodsky:2002ka}, based on the removal of the renormalizaton-scale ambiguity by absorbing the non-conformal $\beta_0$-terms into the running coupling. Since results for Mueller--Navelet jets with BLM optimization proved to be in good agreement with CMS data\tcite{Ducloue:2013bva,Caporale:2014gpa}, the use of this procedure was extended to other semi-hard processes.

The weakness point, however, is that BLM leads to an expansion of energy scales, whose values turn out to be much larger than their natural values (see Section 3.4 of Ref.\tcite{Celiberto:2020wpk}).
Since the sensitivity of both the QCD running coupling and parton densities on large scales is very weak, any kind of study on scale variation around the BLM-prescribed values would be inconclusive.
More importantly, very large scales bring to a substantial reduction of predicted cross sections, thus hampering any possibility of precision studies.
First, encouraging clues that a fair stability under higher-order BFKL corrections has been reached came out just recently\tcite{Celiberto:2020tmb} in the context of the inclusive Higgs-plus-jet production and are expected also in the inclusive Drell--Yan-plus-jet process~\cite{Golec-Biernat:2018kem}. In those cases, the large energy scales that act as stabilizers of rapidity and $p_T$-distributions mostly come from the transverse mass of the emitted boson, while a similar outcome cannot be obtained in the case of light-object emissions.

In this paper, our aim is to study the high-energy behavior of interesting observables for the following semi-hard reaction:
\begin{equation}
 \label{process}
 P(P_1) \, + \, P(P_2) \; \to \; Q{\text{-jet}}(p_Q, y_Q) \, + \, X \, + \, {\rm{jet}}(p_J, y_J) \; .
\end{equation}
Here, a heavy-flavored jet (labeled as $Q$-jet) and a light-flavored jet are emitted in proton-proton collisions with large transverse momenta, $|\vec p_Q|$ and $|\vec p_J|$, and wide separation in rapidity, $\DY = y_Q - y_J$, together with the radiation of an undetected gluon system, $X$.
Various benefits are gained by considering the detection of such a \emph{heavy-light dijet system} (see panel b) of Fig.\tref{fig:process_2p} and Fig.\tref{fig:process_factorization}).
First, light jets can be tagged at larger rapidities than heavy objects. This grants us the possibility of studying our process at large final-state rapidity distances, where a high-energy treatment is motivated. At the same time, statistics for a heavy-plus-light emission is more favorable with respect to the one for a double heavy-flavor detection.
Then, having two jets of distinct kind not only leads to naturally asymmetric kinematic configurations, thus enriching the exclusiveness of the final-state, but also permits us to better focus on and study distinctive traits of the sole heavy-flavored jet.
Finally, in analogy with the outcome of our recent studies on Higgs-plus-jet distributions\tcite{Celiberto:2020tmb}, it gives us a further chance of hunting for effects of stabilization of the BFKL series, eventually provided by the transverse mass of the heavy quark generating the jet. 
One might argue that transverse masses characterizing open-charm or bottom emissions could be not so large. It is worth it to remark, however, that the core contribution to cross sections is at low $p_T$, where the mass of the heavy quark gives an important contribution to energy scales.

The present work lies in the intersection corner between \emph{two directions}. 
On the one hand, it endows semi-hard probes of BFKL with a new choice.
On the other hand, it pushes forward our program on heavy-flavor physics at high energies. Here, the BFKL resummation would serve as a tool i) to investigate heavy-jet distributions covering broader kinematic regimes, where other formalisms are also valid, and ii) to afford a \emph{complementary} description of bound states~\cite{Boussarie:2017oae}, where the main challenge would be embodying in our high-energy theoretical setup the production mechanisms of heavy hadrons.

The structure of this paper is the following. In Section~\ref{theory} we introduce our formal description of the inclusive heavy-light dijet reaction, presenting details regarding impact factors and the way in which the hybrid high-energy collinear factorization is realized, as well as kinematics of the process itself. Then~(Section~\ref{phenomenology}), we discuss our phenomenological analysis on differential distributions. Thus~(Section~\ref{conclusions}), we come out with conclusions and future perspectives.

\section{Inclusive heavy-light dijet production}
\label{theory}

\subsection{LO impact factors}
\label{impact_factors}

Two basic ingredients to construct the cross section for our process are the heavy-quark pair and the light-jet impact factors, the third one being the BFKL Green's function. 
At LO, a heavy-quark pair is produced by a partonic gluon and the $Q$-jet is generated either from the quark or from the antiquark~(Fig.\tref{fig:hqp_IF}). 
The mechanism of production of a heavy-quark jet by a heavy-quark initial state parton has been neglected, since, in the range of $x$ values relevant for this process at the LHC, $10^{-2}$ to $10^{-4}$, and for a value of the factorization scale typical of our study, $\mu_F = 30$ GeV, the charm and bottom distribution functions in the proton are suppressed by factors of $\sim 30$ and $\sim 50$, respectively, with respect to the gluon one.
Conversely, a light jet directly stems from a partonic light quark or from a gluon (Fig.\tref{fig:jet_IF}).

Let us start by considering the upper-vertex impact factor (heavy-quark pair) of our hybrid-factorized cross section~(Fig.\tref{fig:process_factorization}). We introduce the standard Sudakov decomposition for $p_Q$
\begin{equation}
p_Q = z_Q P_1 + \frac{\mQp^2}{z_Q s} P_2 + p_{Q, \perp} \equiv \frac{z_Q}{x_g} p_1 + \frac{x_g}{z_Q} \frac{\mQp^2}{W^2} p_2 + p_{Q, \perp} \;.
\label{SudakovDec1}
\end{equation}
In the first equality, $z_Q$ is the longitudinal momentum fraction of the initial proton $P(P_1)$ carried by the detected heavy quark, $s = 2 P_1 \cdot P_2$ is the hadronic center-of-mass energy, $\vec{p}_Q$ is the transverse momentum of the tagged quark with respect to the proton collision axis, $\mQp = \sqrt{m^2 + \vec{p}_Q^{\; 2}}$ its transverse mass, lastly $p_{Q,\perp} = (0,\vec{p}_Q,0)$. 
The reason for the second equality is to stress that, in Ref.\tcite{Bolognino:2019yls}, the impact factor was calculated by using as light-cone basis the momenta $p_1$ and $p_2$ of the colliding gluons ($W^2 = 2 p_1 \cdot p_2$ is the partonic center-of-mass energy). Hence, when adopting $P_1$ and $P_2$ as light-cone vectors, one must be careful, performing the substitution $z \to \frac{z_Q}{x_g}$, where $x_g$ is the longitudinal fraction of momenta $P_1$ carried by the gluon with momenta $p_1$, and including the right jacobian factor. The LO heavy-quark pair impact factor in the in the $(n, \nu)$-representation, namely after its projection onto the BFKL eigenfunction, turns out to be

\[
\frac{d\Phi_{gg}^{\lbrace{Q\bar{Q}\rbrace}}\left(n,\nu,\vec{p}_Q,z_Q, x_g\right)}{d^2\vec{p}_Q
  \; dz_Q} \equiv \int\frac{d^2\vec{k}}{\pi\sqrt{2}}(\vec{k}^{\;2})^{i\nu-\frac{3}{2}}
e^{in\theta}\frac{d\Phi^{\lbrace{Q\bar{Q}\rbrace}}_{gg}(\vec{k},\vec{p}_Q,z_Q)}{d^2\vec{p}_Q
  \; dz_Q} 
\]
\[
= \frac{\alpha_s^2 \sqrt{N_c^2-1}}{2\pi N_c \; x_g}\left\{ m^2 \left( I_3
- 2\frac{I_2(0)}{M_{Q \perp}^2} \right) + (z^2 + \bar{z}^2)
\left( -m^2 \left(I_3 - 2\frac{I_2(0)}{M_{Q \perp}^2} \right)
+ \frac{I_2(1)}{M_{Q \perp}^2} \right) \right.
\]
\[
\left. - \frac{N_c^2}{N_c^2-1} \Bigg[ 2 m^2 \left[\left( z^2 + \bar{z}^2
 - 1 \right) \left( 1 -  \left(z^2\right)^{\frac{1}{2}-i\nu} \right) \right]
 \frac{I_2(0)}{M_{Q \perp}^2} + \left[ 2m^2(z^2 + \bar{z}^2 -1)
\left(z^2\right)^{\frac{1}{2}-i\nu} \right] 
\right.
\]
\[
\left.
\left( I_3 - \frac{I_4(0)}{\left(z^2\right)^{\frac{1}{2}-i\nu}} \right)
- (z^2 + \bar{z}^2) \bigg[ (1-z)^2 I_4(1) - \frac{\left( 1
    - \left(z^2\right)^{\frac{1}{2}-i\nu} \right)}{M_{Q \perp}^2} I_2(1) \bigg]
\Bigg] \right\} 
\]
\begin{equation}
\equiv \alpha_s^2 \; e^{in \phi_1} c_Q(n, \nu, \vec{p}_Q, z_Q, x_g) \;,
\label{cQ}
\end{equation}
where $z= z_Q / x_g$, $\bar{z}=1-z$, $I_2(\lambda)$, $I_3$ and $I_4(\lambda)$ read
\begin{equation}
\label{AAA}
\begin{split}
I_2\left(\lambda\right)&=\frac{\left(\vec{p}_Q^{\; 2}\right)^{\frac{n}{2}}e^{in\phi_1}}
{\sqrt{2}}\frac{1}{\left(M_{Q \perp}^2\right)^{\frac{3}{2}+\frac{n}{2}-i\nu-\lambda}}
\frac{\Gamma\left(\frac{1}{2}+\frac{n}{2}+i\nu+\lambda\right)
  \Gamma\left(\frac{1}{2}+\frac{n}{2}-i\nu-\lambda\right)}{\Gamma\left(1+n\right)}\\ &\times\frac{\left(\frac{1}{2}+\frac{n}{2}-i\nu-\lambda\right)}
     {\left(-\frac{1}{2}+\frac{n}{2}+i\nu+\lambda\right)}\;
     _2F_1\left(-\frac{1}{2}+\frac{n}{2}+i\nu+\lambda,\frac{3}{2}+\frac{n}{2}
     -i\nu-\lambda,1+n,\zeta\right) \;,
\end{split} 
\end{equation} 
\vspace{0.5 cm}
\begin{equation}
\label{I3_final}
\begin{split}
  I_3= \frac{\left(\vec{p}_Q^{\; 2}\right)^{\frac{n}{2}}e^{in\phi_1}}{\sqrt{2}}
  & \frac{1}{\left(M_{Q \perp}^2\right)^{\frac{5}{2}+\frac{n}{2}-i\nu}}
  \frac{\Gamma\left(\frac{1}{2}+\frac{n}{2}+i\nu\right)
    \Gamma\left(\frac{1}{2}+\frac{n}{2}-i\nu\right)}
       {\Gamma\left(1+n\right)}\frac{\left(\frac{1}{2}+\frac{n}{2}-i\nu\right)}
       {\left(-\frac{1}{2}+\frac{n}{2}+i\nu\right)}\;
       \\ & \times \left(\frac{3}{2}+\frac{n}{2}-i\nu\right)\;
       _2F_1\left(-\frac{1}{2}+\frac{n}{2}+i\nu,\frac{5}{2}+\frac{n}{2}
       -i\nu,1+n,\zeta \right) \;,
\end{split} 
\end{equation}
\vspace{0.5 cm}
\begin{equation}
\begin{split}
  I_4\left(\lambda\right) = & \frac{\left(\vec{p}_Q^{\; 2}\right)^{\frac{n}{2}}
    e^{in\phi_1}}{z^2\sqrt{2}}  \frac{\left(\frac{3}{2}-i\nu-\lambda
+\frac{n}{2}\right)}{\left(M_{Q \perp}^2\right)^{\frac{5}{2}-i\nu-\lambda+\frac{n}{2}}} \frac{\Gamma\left(\frac{1}{2}+\frac{n}{2}+i\nu+\lambda \right)\Gamma\left(\frac{1}{2}+\frac{n}{2}-i\nu-\lambda \right)}{\Gamma\left(1+n\right)} \\ & \times \frac{\left(\frac{1}{2}+\frac{n}{2}-i\nu-\lambda \right)}{\left(-\frac{1}{2}+\frac{n}{2}+i\nu+\lambda \right)} \int_0^1 d\Delta \left(1+\frac{\Delta}{z}-\Delta\right)^n \left(1+\frac{\Delta}{z^2}-\Delta\right)^{-\frac{5}{2}+i\nu+\lambda-\frac{n}{2}} \; \\ & \times \; _2F_1\left(-\frac{1}{2}+i\nu+\lambda+\frac{n}{2},\frac{5}{2}-i\nu-\lambda+\frac{n}{2},1+n,\zeta \;  \frac{\left(1+\frac{\Delta}{z}-\Delta\right)^2}{\left(1+\frac{\Delta}{z^2}-\Delta\right)}\right) \;,
\end{split}
\label{I4}
\end{equation}
and $\zeta \equiv \frac{\vec{p}_Q^{\; 2}}{\mQp^2}$; the azimuthal
angles $\theta$, $\phi_1$ are defined as $\cos \theta \equiv k_{x}/|\vec{k}|$ and $\cos \phi_1 \equiv p_{Q,x}/|\vec{p}_Q|$. \\
Let us now turn our attention on the lower-vertex~(Fig.\tref{fig:process_factorization}) impact factor (light jet). We apply the standard Sudakov decomposition for $p_J$, having so
\begin{equation}
p_J = z_J P_2 + \frac{|\vec{p}_J|^2}{z_J s} P_1 + p_{J \perp} \;,
\label{SudakovDec2}
\end{equation}
where $z_J$ is the longitudinal fraction of momenta of the initial proton $P(P_2)$ carried by the jet and $\vec{p}_J$ its transverse momenta with respect to the collision axis.
The expression for the LO impact factor depends on which type of parton starts the process. In the $(n, \nu)$-representation we have:
\begin{equation*}
    \left( \frac{d\Phi_{qq}^{\lbrace{q\rbrace}}\left(n,\nu,\vec{p}_J,z_J\right)}{d^2\vec{p}_J
  \; dz_J} \right)^* = \frac{C_F}{C_A} \left( \frac{d\Phi_{gg}^{\lbrace{g\rbrace}}\left(n,\nu,\vec{p}_J,z_J\right)}{d^2\vec{p}_J
  \; dz_J} \right)^* = \int\frac{d^2\vec{k}}{\pi\sqrt{2}}(\vec{k}^{\;2})^{-i\nu-\frac{3}{2}}
e^{-in\theta}\frac{d\Phi^{\lbrace{q\rbrace}}_{qq}(-\vec{k},\vec{p}_J,z_J)}{d^2\vec{p}_J
  \; dz_J} 
  \end{equation*}
  \begin{equation}
  = \alpha_s e^{- i n (\phi_2 + \pi)} 2 \sqrt{\frac{C_F}{C_A}} (\vec{p}_J^{\; 2})^{- i \nu - 3/2} \delta (x_J - z_J)  \equiv \alpha_s e^{- i n (\phi_2 + \pi)} \left[ c_J (n , \nu , |\vec{p}_J|) \right]^{*} \delta (x_J - z_J) \;,
\end{equation}
where $d\Phi_{qq}^{\lbrace{q\rbrace}}$ and $d\Phi_{gg}^{\lbrace{g\rbrace}}$ are the quark and gluon LO impact factors, respectively, $x_J$ is the longitudinal fraction of momenta of the initial proton $P(P_2)$ carried by the parton entering the jet vertex. The azimuthal angle $\phi_2$ is defined $\cos \phi_2 \equiv p_{J,x}/ |\vec{p}_J|$.
Since our impact factors are taken at LO, the jet selection function is trivial and merely identifies the kinematics of the produced parton with that of the jet.

\subsection{Kinematics of the process}
\label{kinematics}
Using the Sudakov decomposition (\ref{SudakovDec1}) and denoting with $p_Q = (E_Q, \vec{p}_Q, p_{Q||})$, we can express the rapidity of the tagged quark as
\begin{equation}
    y_Q = \frac{1}{2} \ln \left( \frac{E_Q + p_{Q||}}{E_Q - p_{Q||}} \right) = \ln \left( \frac{2 z_Q E_{p_1}}{\mQp} \right) \;,
\end{equation}
where $P_1 = E_{p_1}(1,\vec{0},1)$ and $P_2 = E_{p_2}(1,\vec{0},-1)$, and hence $s = 4 E_{p_1} E_{p_2}$.
Using the Sudakov decompositions (\ref{SudakovDec2}) and denoting with $p_J = (E_J, \vec{p}_J, p_{J||})$ we can express the rapidity of the jet as
\begin{equation}
    y_J = - \ln \left( \frac{2 x_J E_{p_2}}{|\vec{p}_J|} \right) \;.
\end{equation}
The difference of rapidity is
\begin{equation}
    \DY \equiv y_Q - y_J = \ln \left( \frac{z_Q \; x_J \; s}{|\vec{p}_J| \mQp} \right) 
\end{equation}
and the semi-hard kinematic requirement imposes
\begin{equation}
   \frac{s}{|\vec{p}_J| \mQp} = \frac{e^{\DY}}{z_Q  x_J} \gg 1 \;.
\end{equation}
The Jacobian of the transformation from longitudinal momentum fractions to final-state ra\-pi\-di\-ty is
\begin{equation}
\drv z_Q \drv x_J = \frac{e^{\DY}}{s} |\vec{p}_J| \mQp \drv y_Q \drv y_J \;.
\end{equation}

\subsection{Proton-proton cross section}
\label{cross_section}
In order to pass from the hard subprocess to the physical one, initiated by proton-proton collisions~(Fig.\tref{fig:process_factorization}), we include the contribution of the partons distribution inside the two colliding particles.
Then, the differential proton cross section can be expressed as
\begin{equation}
\label{crosfin}
\frac{d\sigma_{pp}}{\drv y_Q \drv y_J \drv |\vec{p}_Q| \drv |\vec{p}_J| \drv \phi_1 \drv \phi_2}=\frac{1}{(2\pi)^2} \left[\mathcal{C}_0+2 \sum_{n=1}^{\infty} \cos(n\phi) \mathcal{C}_n \right], 
\end{equation}
where $\phi = \phi_1 - \phi_2 - \pi$ and with the azimuthal coefficients
\[
 \mathcal{C}_n = \frac{e^{\DY} |\vec{p}_Q| |\vec{p}_J|^2 \mQp }{s} \int_{x_g^{\rm min}}^1 \drv x_g f_g(x_g, \mu_{F_1}) \tilde{f} (x_J, \mu_{F_2})
\]
\begin{equation}
 \int_{-\infty}^{+\infty} \drv \nu \left(\frac{W^2}{s_0}\right)^{\bar{\alpha}_s\left(\mu_R\right)\chi\left(n,\nu\right)+\bar{\alpha}_s^2\left(\mu_R\right)\left[\bar{\chi}\left(n,\nu \right)+\frac{\beta_0}{8N_c}\chi\left(n,\nu\right)\left(-\chi \left(n,\nu\right)+\frac{10}{3}+2 \ln\frac{\mu_R^2}{\mQp \vec p_J} \right)\right]}
\label{Cn}
\end{equation}
\[
 \times\; \alpha_s^3 \left(\mu_{R}\right) 
 c_Q\left(n,\nu,\vec{p}_Q, z_{Q},x_g \right) 
 [c_J\left(n,\nu,\vec{p}_J \right)]^{*}
\]
\[
 \times\; \left\{ 1 + \frac{c_Q^{(1)} \left(n,\nu,\vec{p}_Q, z_{Q} \right)}{c_Q \left(n,\nu,\vec{p}_Q, z_{Q} \right)} + \left[ \frac{c_J^{(1)} \left(n,\nu,\vec{p}_J,x_J \right)}{c_J \left(n,\nu,\vec{p}_J \right)} \right]^{*} 
 + \bar{\alpha}_s^2 \left(\mu_{R}\right) \ln \left(\frac{W^2}{s_0}\right) \frac{\beta_0}{4 N_c} \chi \left(n, \nu \right) f_Q\left(\nu\right) \right\} \; .
\]
where the lower bound in the $x_g$-integration follows from kinematics, $x_g^{\rm min} = e^{-(y_Q^{\rm max}-y_Q)}$,
\begin{equation}
\label{fnu}
    f_Q(\nu) = \frac{i}{2} \frac{\drv}{\drv \nu} \ln c_Q + \ln \mQp \; ,
\end{equation}
$\mu_{F_{1,2}}$ are the factorization scales and, for the sake of convenience, we define an effective collinear PDF as
\begin{equation}
 \tilde{f} (x_J, \mu_{F_2}) = \frac{C_A}{C_F} f_g(x_J, \mu_{F_2}) + \sum_{a=q,\bar{q}} f_a (x_J, \mu_{F_2}) \; .  
\end{equation}
In Eq.~(\ref{Cn}), $N_c$ is the color number, $\mu_R$ and $\mu_{F_{1,2}}$ are renormalization and factorization scales, $\bar{\alpha}_s \equiv N_c  \alpha_s / \pi $, 
\begin{equation}
    \chi (n,\nu) = 2 \psi(1) - \psi(\frac{n}{2} + \frac{1}{2} + i \nu) - \psi(\frac{n}{2} + \frac{1}{2} - i \nu)
\end{equation}
are the LO BFKL eigenfunctions,
\begin{equation}
    \beta_0 = \frac{11 N_c}{3} - \frac{2 n_f}{3}
\end{equation}
is the first coefficients of the QCD $\beta$-function, while $c_Q^{(1)} \left(n,\nu,\vec{p}_Q, z_{Q} \right)$ and $c_J^{(1)} \left(n,\nu,\vec{p}_J ,x_J \right)$ are NLO universal correction to the heavy quark and jet impact factors, respectively.  \\ 
The NLO correction to the heavy-quark pair impact factor reads
\[
  \frac{c_Q^{(1)} \left(n,\nu,\vec{p}_Q, z_{Q} \right)}{c_Q \left(n,\nu,\vec{p}_Q, z_{Q} \right)} = \bar{\alpha}_s \left(\mu_{R}\right) \frac{\chi(n, \nu)}{2} \ln \left(\frac{s_0}{\mQp^2} \right) 
\]
\begin{equation}
\label{cQ1}
  + \; \bar{\alpha}_s \left(\mu_{R}\right) \frac{\beta_0}{4 N_c} \left(\frac{5}{3} + 2 \ln \frac{\mu_{R}^2}{\mQp |\vec{p}_Q|} + 2 f_Q\left(\nu\right) \right) 
\end{equation}
\[
  - \; \frac{1}{f_g (x_g, \mu_{F_1})} \ln \left( \frac{\mu_{F_1}^2}{\mQp^2} \right) \frac{\bar{\alpha}_s (\mu_{F_1})}{2 N_c} \int_{x_g}^{1} \frac{\drv z}{z} \left( P_{gg} (z) f_{g} \left( \frac{x_g}{z} , \mu_{F_1} \right) + \sum_{a = q, \bar{q}} P_{ga} (z) f_{a} \left( \frac{x_g}{z} , \mu_{F_1} \right) \right) \; , 
\]
whereas the NLO correction to the light-jet impact factor is
\[
   \left[ \frac{c_J^{(1)} \left(n,\nu,\vec{p}_J ,x_J \right)}{c_J \left(n,\nu,\vec{p}_J \right)} \right]^{*} = \bar{\alpha}_s \left(\mu_{R}\right) \frac{\chi(n, \nu)}{2} \ln \left(\frac{s_0}{|\vec{p}_J|^2} \right) + \bar{\alpha}_s \left(\mu_{R}\right) \frac{\beta_0}{4 N_c} \left( 2 \ln \frac{\mu_{R}}{|\vec{p}_J|} + \frac{5}{3} \right) 
\]
\[
   - \; \ln \left( \frac{\mu_{F_2}^2}{|\vec{p}_J|^2} \right)  \frac{ \bar \alpha_s(\mu_{F_2}) }{2 N_c \tilde{f} (x_J, \mu_{F_2})} \left[ \frac{C_A}{C_F} \int_{x_J}^1 \frac{\drv z}{z} \left( P_{gg} (z) f_{g} \left( \frac{x_J}{z} , \mu_{F_2} \right) + \sum_{a = q, \bar{q}} P_{ga} (z) f_{a} \left( \frac{x_J}{z} , \mu_{F_2} \right) \right) \right.
\]
\begin{equation}
\label{cJ1}
   \left. + \; \sum_{a = q, \bar{q}} \int_{x_J}^1 \frac{\drv z}{z} \left( P_{ag}(z) f_g \left( \frac{x_J}{z}, \mu_{F_2} \right) + P_{aa}(z) f_a \left( \frac{x_J}{z}, \mu_{F_2} \right) \right) \right] \; .
\end{equation}

\section{Phenomenology}
\label{phenomenology}

\subsection{Observables and final-state configurations}
\label{observables}

We include in our phenomenological analysis three classes of observables:

\begin{itemize}

    \item \emph{Azimuthal-angle coefficients}, integrated over the phase space of the outgoing jets, at fixed values of their mutual rapidity separation, $\DY$. One has
    \begin{equation}
     \label{Cn_int}
     C_n(\DY, s) =
     \int_{p^{\rm min}_Q}^{p^{\rm max}_Q} \drv |\vec p_Q|
     \int_{p^{\rm min}_J}^{{p^{\rm max}_J}} \drv |\vec p_J|
     \int_{y^{\rm min}_Q}^{y^{\rm max}_Q} \drv y_Q
     \int_{y^{\rm min}_J}^{y^{\rm max}_J} \drv y_J
     \, \delta \left( y_Q - y_J - \DY \right)
     \, {\cal C}_n 
     \, ,
    \end{equation}
    where $C_0$ represents the $\varphi$-summed cross section differential in $\DY$ (\emph{alias}, $\DY$-distribution).
    We go with realistic kinematic configurations, typical of current and forthcoming analyses at the LHC, by allowing the transverse momentum of the $Q$-jet, $|\vec p_Q|$, to be in the range from 20 GeV to 100 GeV (typical of the bottom-jet detection at CMS\tcite{Chatrchyan:2012dk,Chatrchyan:2012jua}), while the light jet transverse momentum, $|\vec p_J|$, ranges from 20 GeV to 60 GeV\tcite{Khachatryan:2016udy}. We investigate the $\DY$-behavior of the $\varphi$-summed cross section (or $\DY$-distribution), $C_0(\DY, s)$, of the azimu\-thal-correlation moments, $R_{n0}(\DY, s) = C_{n}/C_{0} \equiv \langle \cos n \varphi \rangle$, and of their ratios, $R_{nm} = C_{n}/C_{m}$~\cite{Vera:2006un,Vera:2007kn}.

    \item \emph{Azimuthal distribution} of the two tagged jets, as a function of $\varphi$ and at fixed values of $\DY$, given as
    \begin{equation}
     \label{azimuthal_distribution}
     \frac{\drv \sigma_{pp} (\varphi, \DY, s)}{\sigma_{pp} \drv \varphi} = \frac{1}{\pi} \left\{ \frac{1}{2} + \sum_{n=1}^{\infty} \cos(n \varphi) \langle \cos(n \varphi) \rangle \right\}
     \equiv \frac{1}{\pi} \left\{ \frac{1}{2} + \sum_{n=1}^{\infty} \cos(n \varphi) R_{n0} \right\} \; .
    \end{equation}
    This distribution represents one of the most directly accessible observables in experimental analyses. 
    Indeed, experimental measurements hardly cover the whole azimuthal-angle plane due to limitations of the apparatus. Therefore, distributions differential on the azimuthal-angle difference, $\varphi$, could be easier compared with data. At the same time, technical difficulties arise in the numeric calculation of Eq.~(\ref{azimuthal_distribution}) since several $C_n$ coefficients need to be calculated, with instabilities in the $\nu$-integration emerging when $n$ grows.
    Ranges for $|\vec p_Q|$ and $|\vec p_J|$ are the ones listed in the previous point.

    \item \emph{Transverse-momentum distribution} of the $Q$-jet at fixed values of $\DY$ (\emph{alias}, $p_Q$-di\-stri\-bu\-tion),
    \begin{equation}
     \label{pT_distribution}
     \frac{\drv \sigma_{pp}(|\vec p_Q|, \DY, s)}{\drv |\vec p_Q| \drv \DY} =
     \int_{p^{\rm min}_J}^{{p^{\rm max}_J}} \drv |\vec p_J|
     \int_{y^{\rm min}_Q}^{y^{\rm max}_Q} \drv y_Q
     \int_{y^{\rm min}_J}^{y^{\rm max}_J} \drv y_J
     \, \delta \left( y_Q - y_J - \DY \right)
     \, {\cal C}_0 
     \, ,
    \end{equation}
    while the light-jet transverse momentum is in the range 35 GeV $< |\vec p_J| <$ 60 GeV.

\end{itemize}

In all the considered cases, we allow for light-jet detection both in the CMS barrel and in the endcaps, having so $|y_J| < 4.7$.
As for the $Q$-jet rapidity, we consider a barrel rapidity-range, $|y_Q| < 2.5$.

We stress that our theoretical setup corresponds to an inclusive final state with a heavy and a light jet detected in a given kinematics and with a given rapidity separation, no matter if other (heavy or light) jets are present and no matter what their rapidity separation is. This implies that a single experimental event with a final state featuring more than two hard jets  selected (at least one of them being heavy) should contribute to multiple bins of the $\Delta Y$-distribution, one bin for each pair made by one heavy and one light jet taken in such multi-jet event.

By accounting for the tag of the light jet not only in CMS but also in the CASTOR ultra-backward detector ($-6.6 < y_J < -5.2$)\tcite{Baur:2019yfg,Khachatryan:2020mpd}, one could consider even larger rapidity distances, up to~$\DY~\simeq~9$. It has been recently pointed out\tcite{Celiberto:2020rxb}, however, that in this kinematics large values of partons' $x$
effectively restrict the weight of the undetected gluon radiation. This leads, when inclusive observables are considered, to an incomplete cancellation between virtual and real gluon-emission terms, which results in the appearance of large Sudakov-type double logarithms (threshold double logarithms)\tcite{Bonciani:2003nt,deFlorian:2005fzc,Muselli:2017bad} in the perturbative series, that have to be resummed to all orders. This resummation has not been yet encoded in our formalism. Therefore, we postpone the investigation of our process in CASTOR-jet configurations to future, dedicated analyses.

In our analysis we consider two possible kinds of heavy-flavored emissions: bottom-jets ($b\text{-jets}$, with $m~\equiv~m_b~=~4.18$ GeV/c$^2$) and charm-jets ($c\text{-jets}$, with $m~\equiv~m_c~=~1.2$ GeV/c$^2$).
For the sake of simplicity, we will refer to these two cases as $b\text{-jet}$ and $c\text{-jet}$ channel, respectively.
The center-of-mass energy is fixed at $\sqrt{s} = 14$ TeV.
All calculations are done in the ${\overline {\rm MS}}$ scheme.

\subsection{Numerical technology and uncertainty estimate}
\label{numerics}

All the numerical results were obtained via \textsc{Jethad}\tcite{Celiberto:2020wpk}, a \textsc{Fortran2008-Python3} hybrid interface under development at our Group.
Quark and gluon PDFs were calculated through the \textsc{MMHT2014} NLO PDF
parameterization~\cite{Harland-Lang:2014zoa} as provided by the LHAPDFv6.2.1
library~\cite{Buckley:2014ana}, while we set a two-loop running coupling setup with $\alpha_s\left(M_Z\right)=0.11707$.

The two primary sources of numerical uncertainty arise from the four-dimensional integration over the final-state phase space~(Eq.~(\ref{Cn_int})), together with the integration over~$\nu$~(Eq.~(\ref{Cn})) and from the one-dimensional integral over the longitudinal momentum fraction, $\zeta$, in the NLO corrections to the two jet impact factors (Eqs.~(\ref{cQ1}) and~(\ref{cJ1})).
They were directly estimated by the \textsc{Jethad} parallel integration routines.
Other potential uncertainties, like the upper cutoff in the numerical integration over the $\nu$-variable in~Eq.~(\ref{Cn}) and over $\Delta$-integration in~Eq.~(\ref{cQ}), turned out to be negligible in comparison the first ones.

We weighed the effect of simultaneously varying the renormalization scale, $\mu_R$, and the factorization ones, $\mu_{F_1}$ and $\mu_{F_2}$,
around their \emph{natural} values in the range from 1/2 to two. 
The parameter $C_{\mu}$ entering the inset of panels in Section~\ref{results} expresses the ratio
\begin{equation}
 \label{C_mu}
 C_\mu =
 \frac{\mu_R}{\sqrt{M_{Q,\perp} |\vec p_J|}} =
 \frac{\mu_{{F}_1}}{M_{Q,\perp}} = 
 \frac{\mu_{{F}_2}}{|\vec p_J|}
  \; .
\end{equation}

\subsection{Results and discussion}
\label{results}

\begin{figure}[t]
\centering
\includegraphics[scale=0.55,clip]{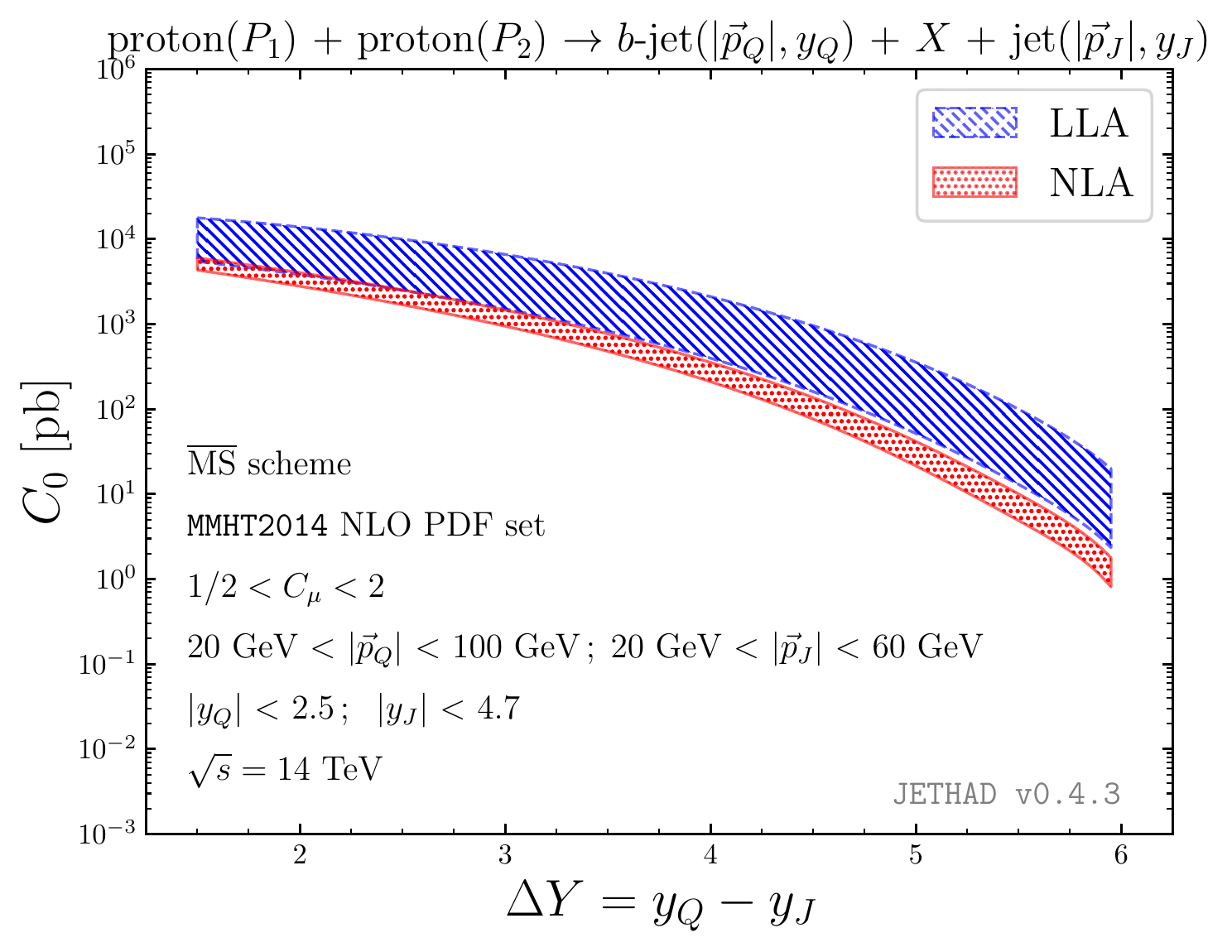}
\includegraphics[scale=0.55,clip]{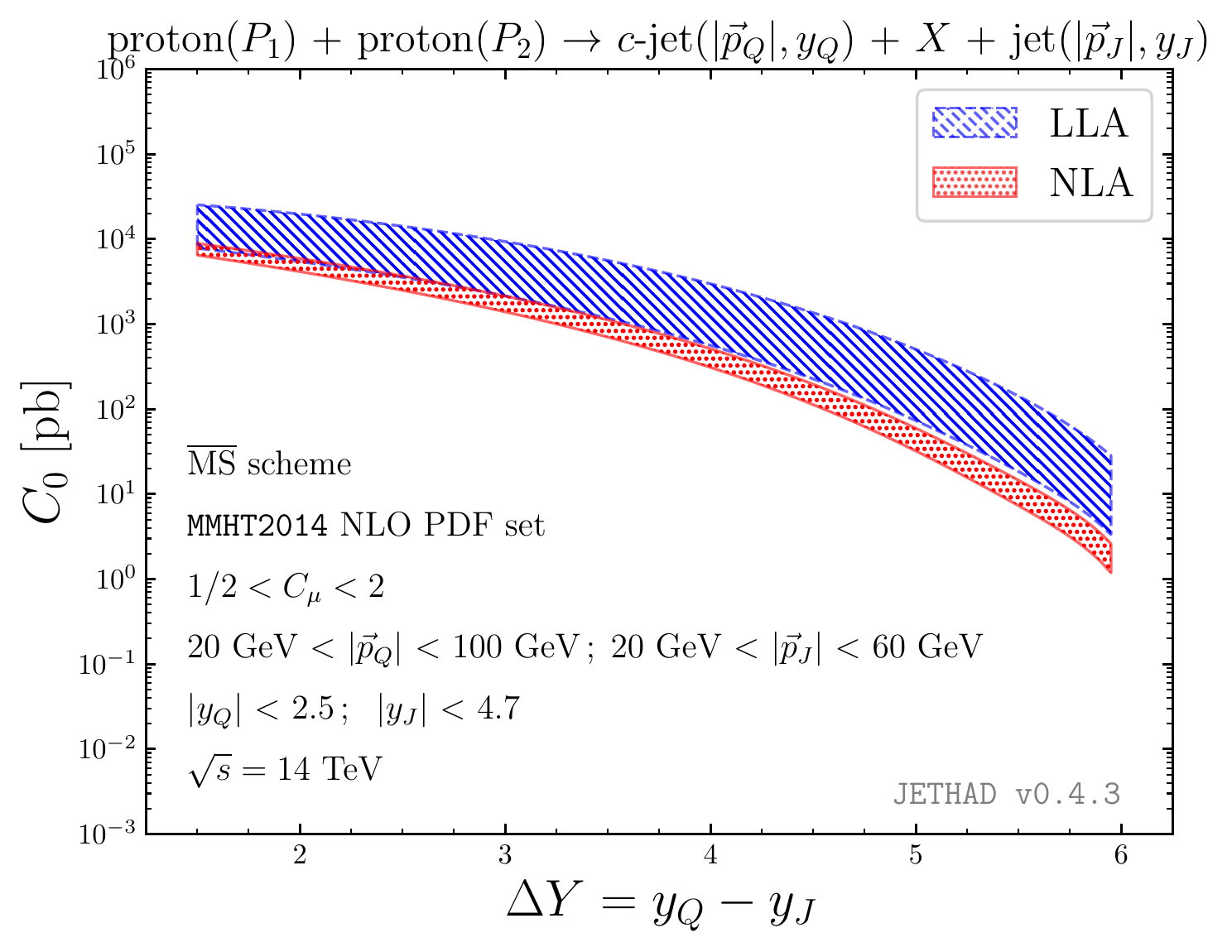}
\caption{$\Delta Y$-dependence of the $\varphi$-summed cross section, $C_0$, in the $b\text{-jet}$ (left) and $c\text{-jet}$ (right) channels and for $\sqrt{s} = 14$ TeV. Shaded bands provide with the combined uncertainty coming from scale variation and numerical integration(s).}
\label{fig:C0}
\end{figure}

\begin{figure}[p]
\centering
\includegraphics[scale=0.54,clip]{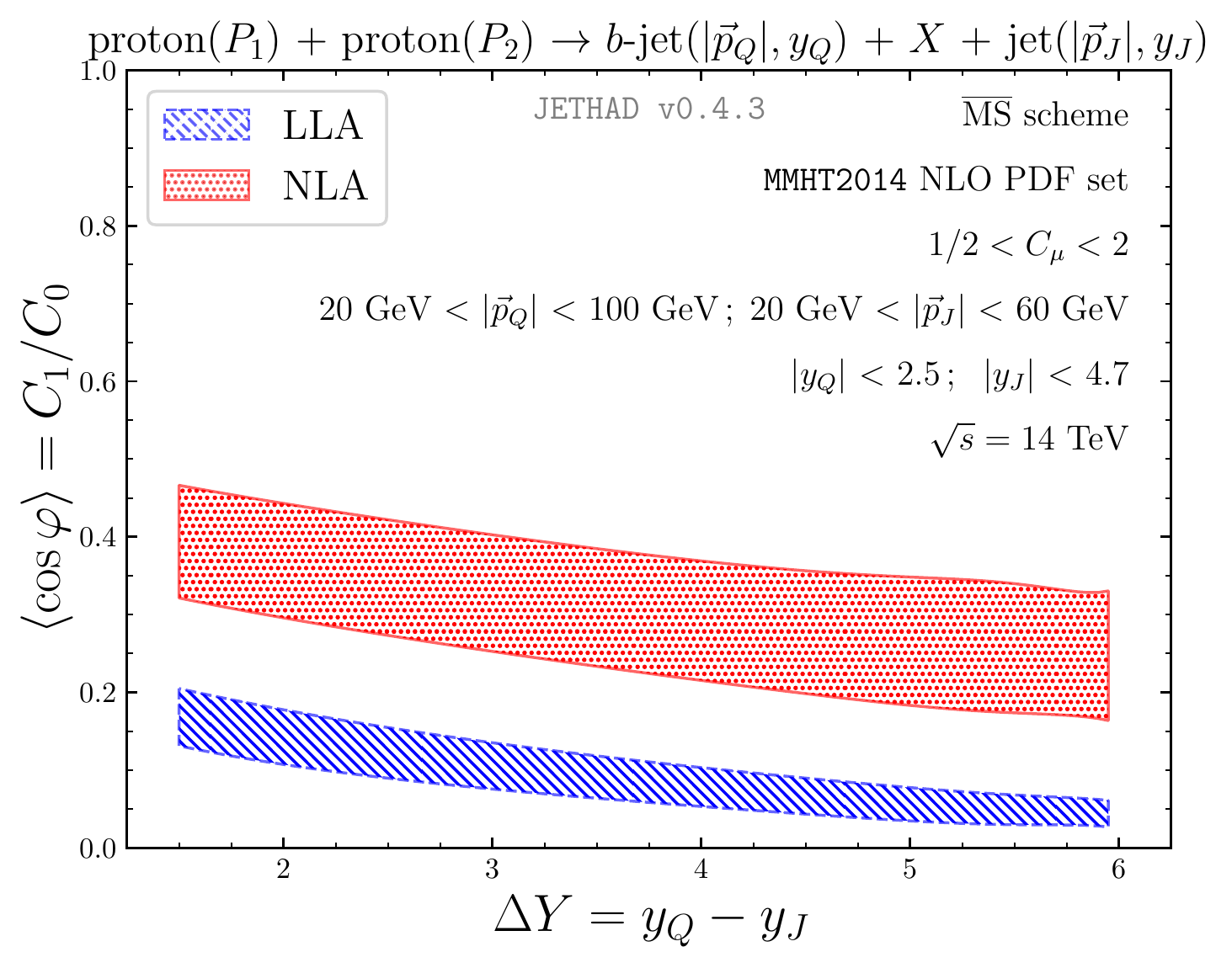}
\includegraphics[scale=0.54,clip]{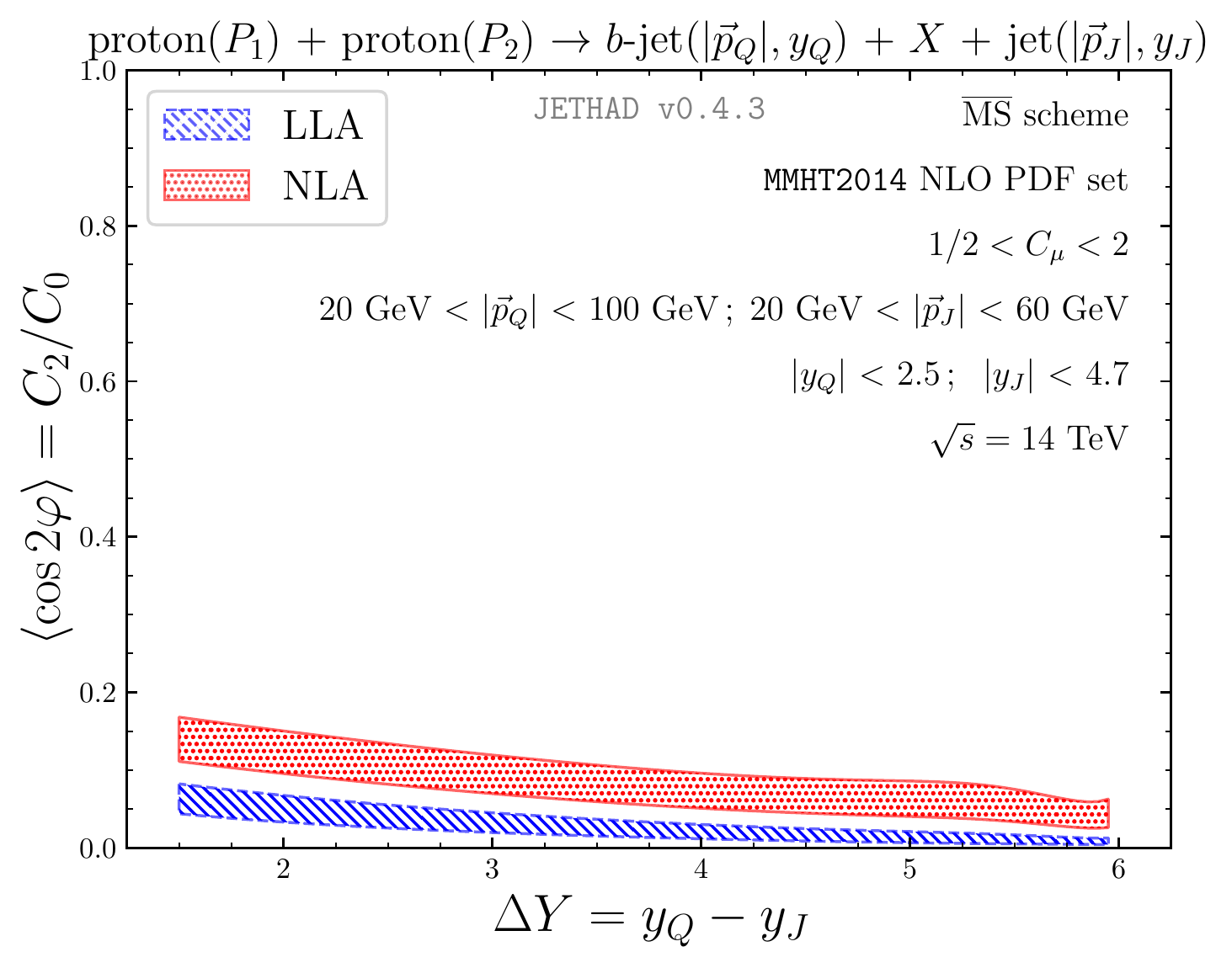}

\includegraphics[scale=0.54,clip]{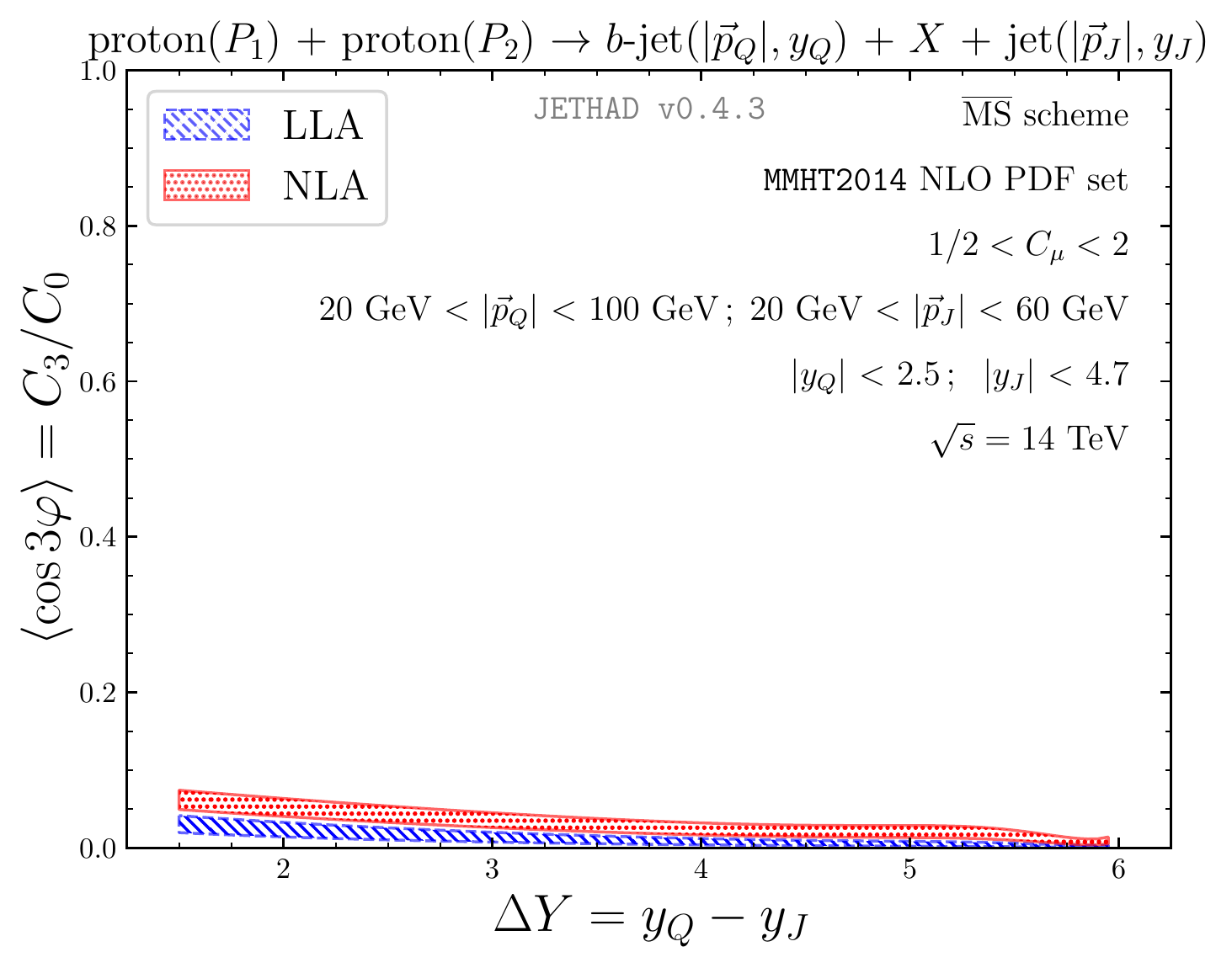}
\includegraphics[scale=0.54,clip]{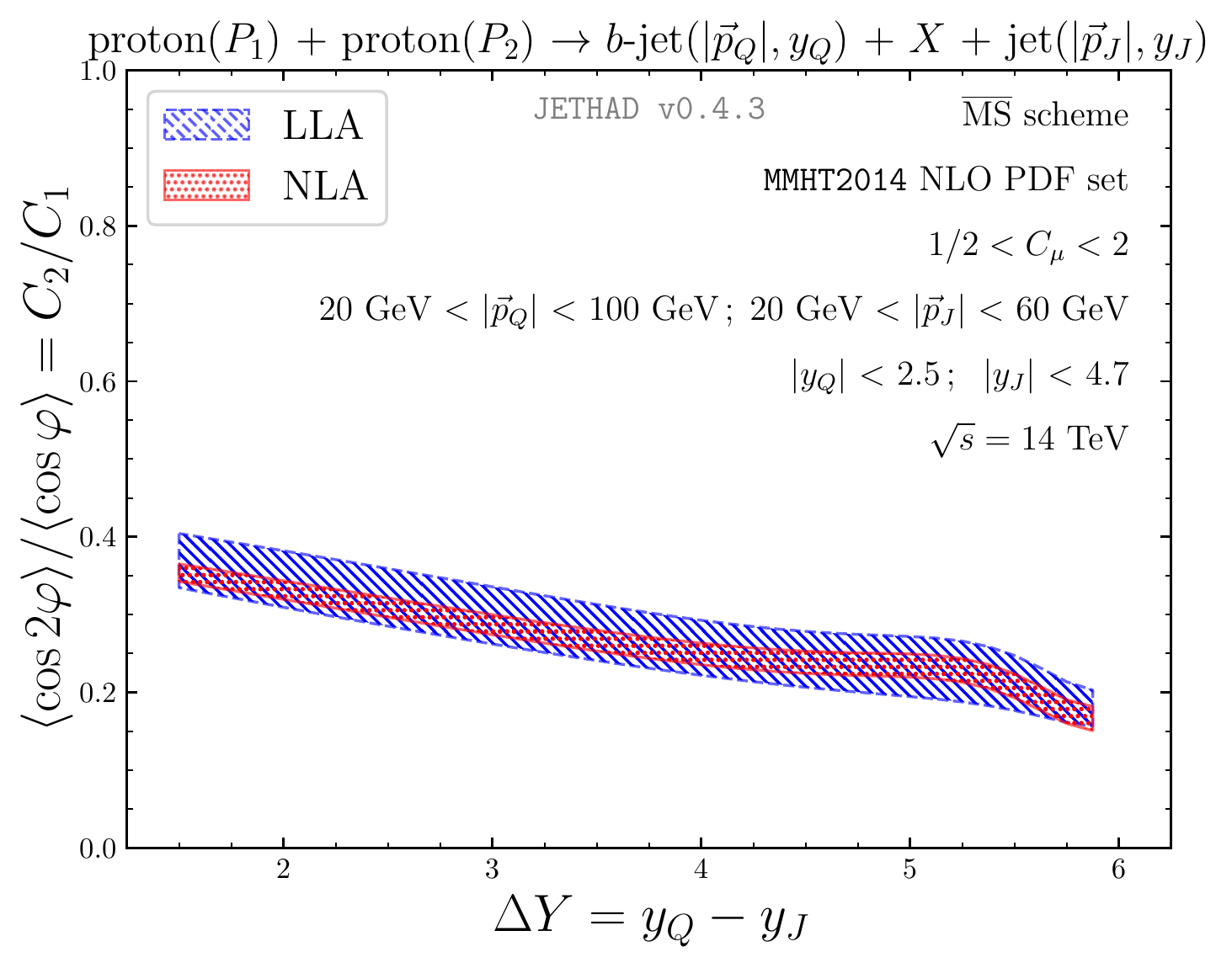}
\caption{$\Delta Y$-dependence of $R_{nm} \equiv C_{n}/C_{m}$ ratios, in the $b\text{-jet}$ channel and for $\sqrt{s} = 14$ TeV. Shaded bands provide with the combined uncertainty coming from scale variation and numerical integration(s).}
\label{fig:Rnm_b}
\end{figure}

\begin{figure}[p]
\centering
\includegraphics[scale=0.54,clip]{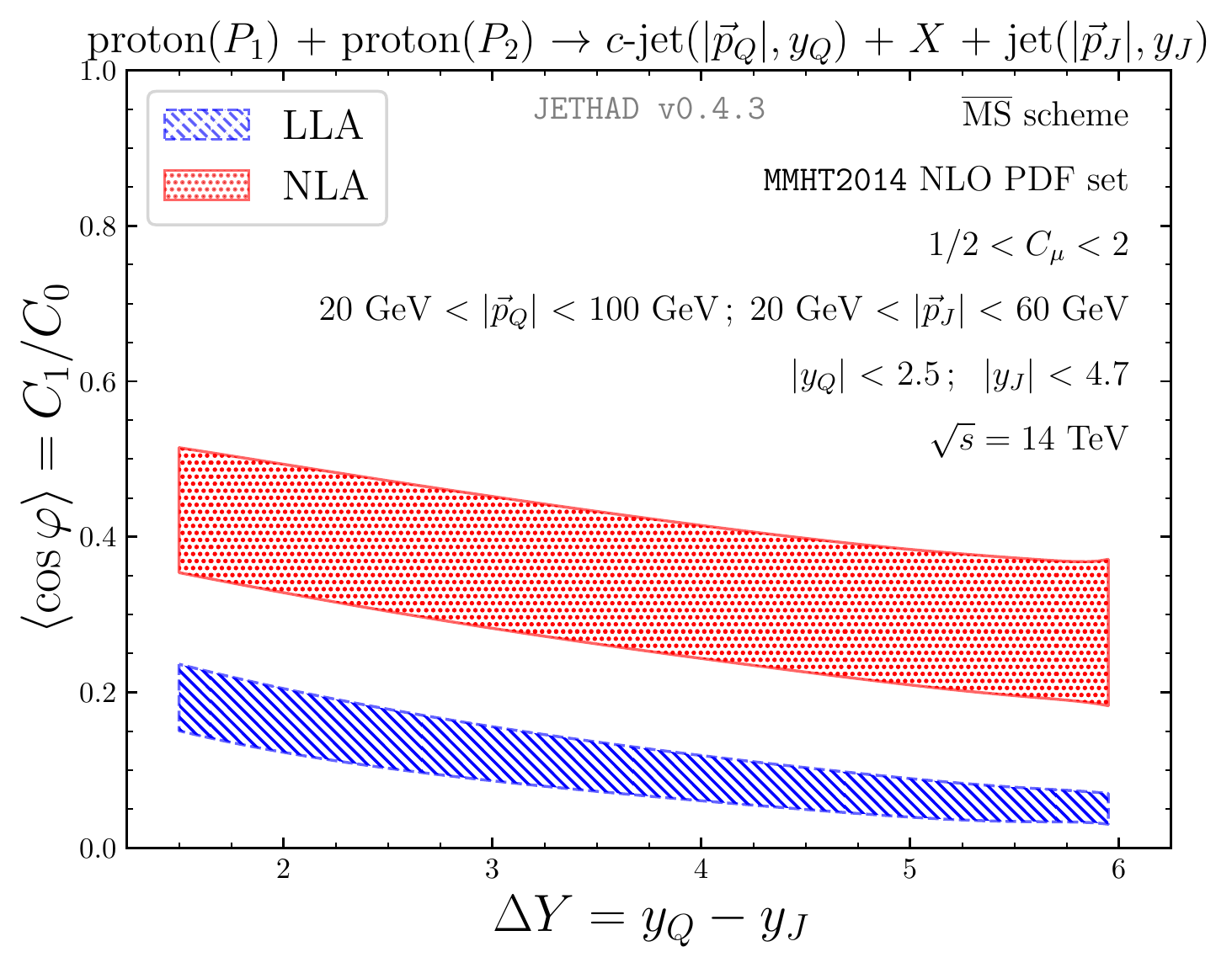}
\includegraphics[scale=0.54,clip]{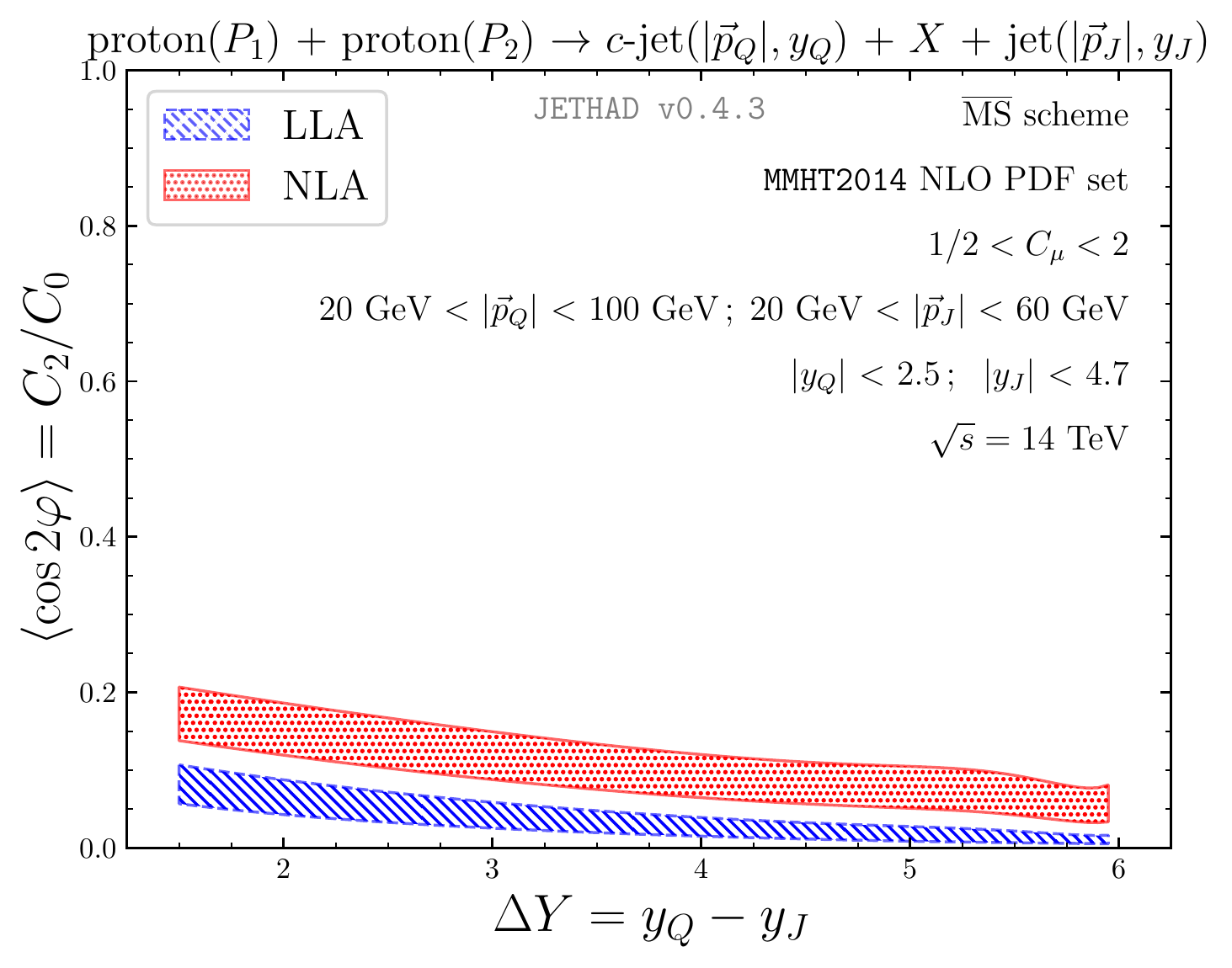}

\includegraphics[scale=0.54,clip]{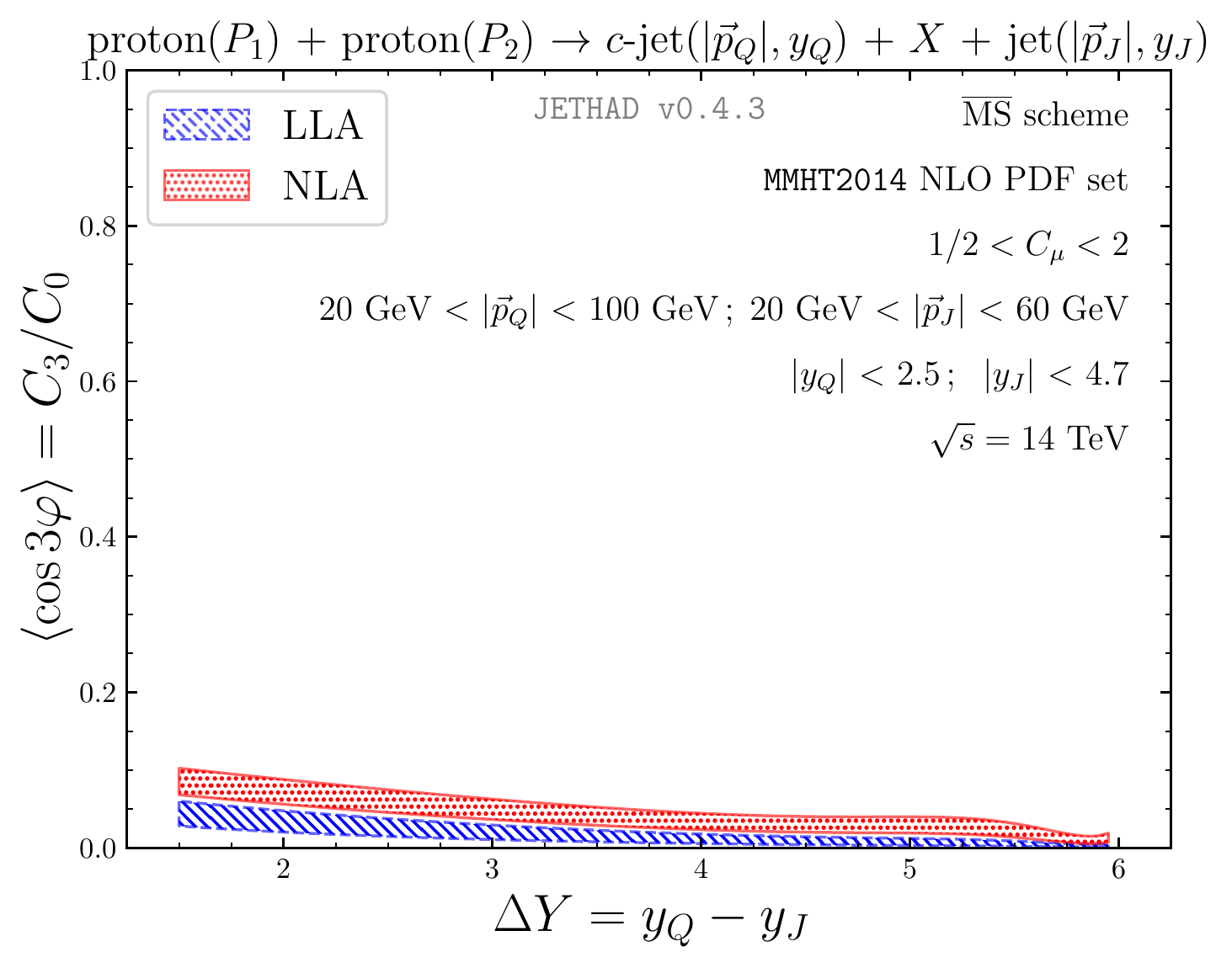}
\includegraphics[scale=0.54,clip]{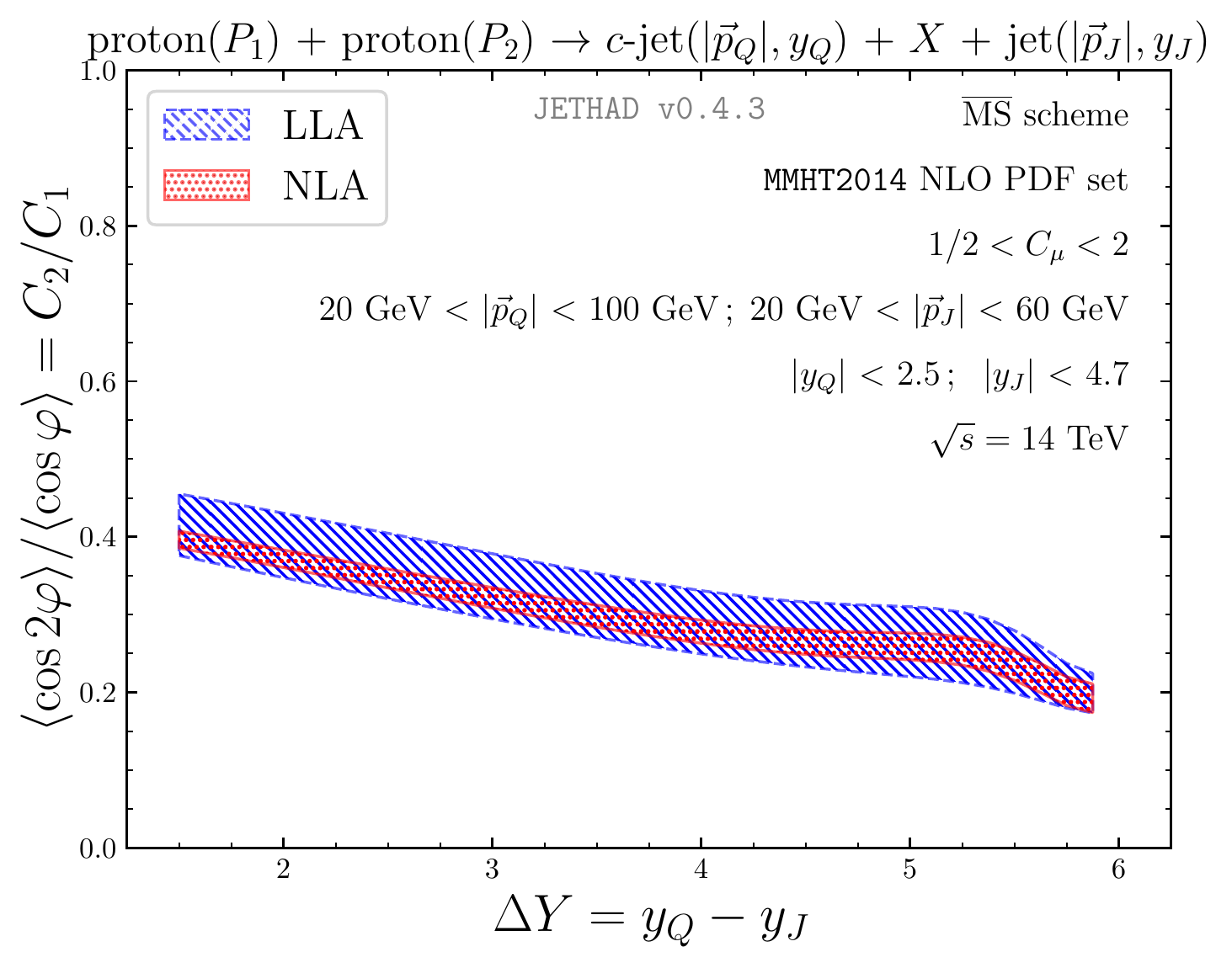}
\caption{$\Delta Y$-dependence of $R_{nm} \equiv C_{n}/C_{m}$ ratios, in the $c\text{-jet}$ channel and for $\sqrt{s} = 14$ TeV. Shaded bands provide with the combined uncertainty coming from scale variation and numerical integration(s).}
\label{fig:Rnm_c}
\end{figure}

\begin{figure}[p]
\centering
\includegraphics[scale=0.53,clip]{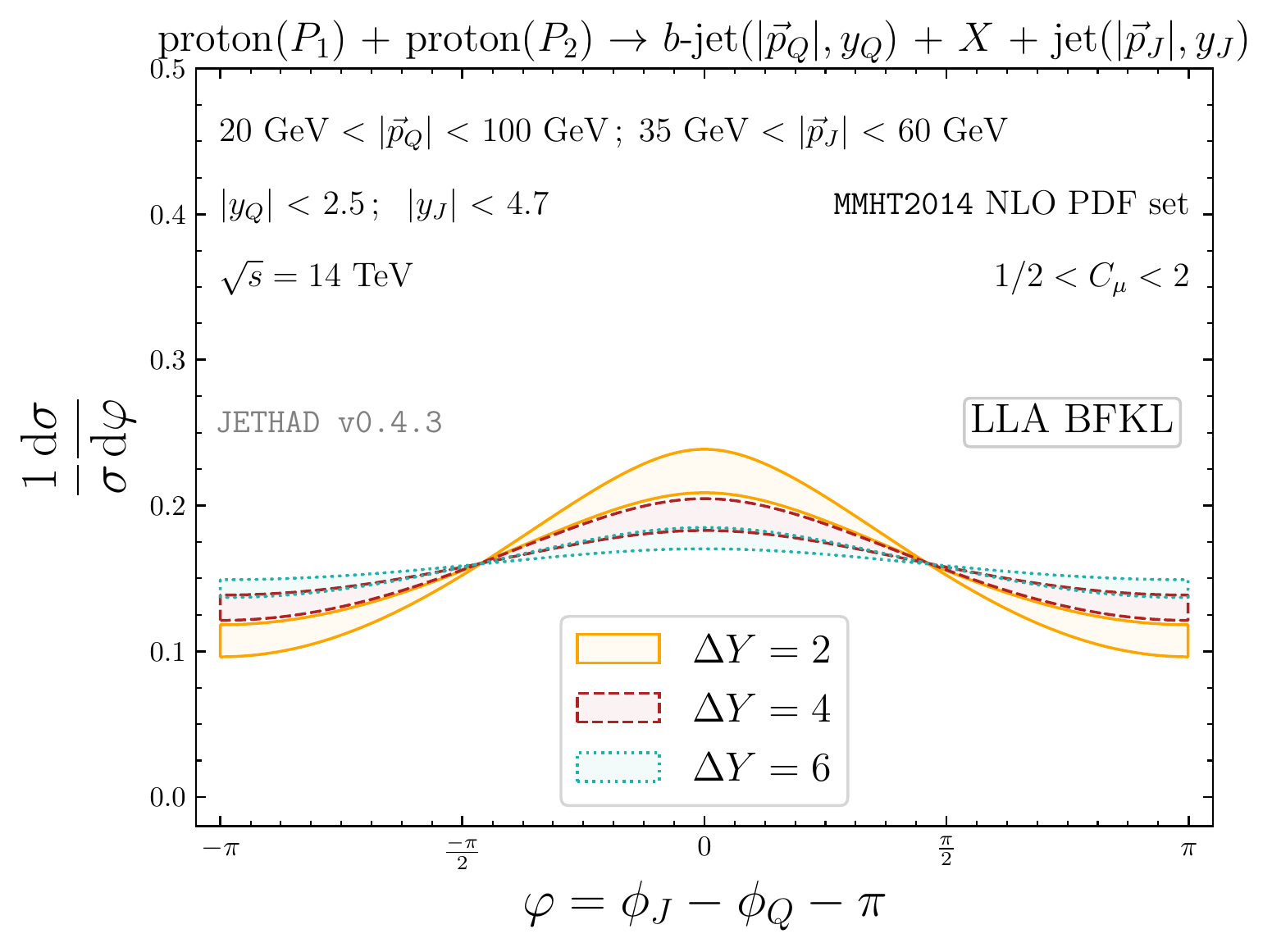}
\includegraphics[scale=0.53,clip]{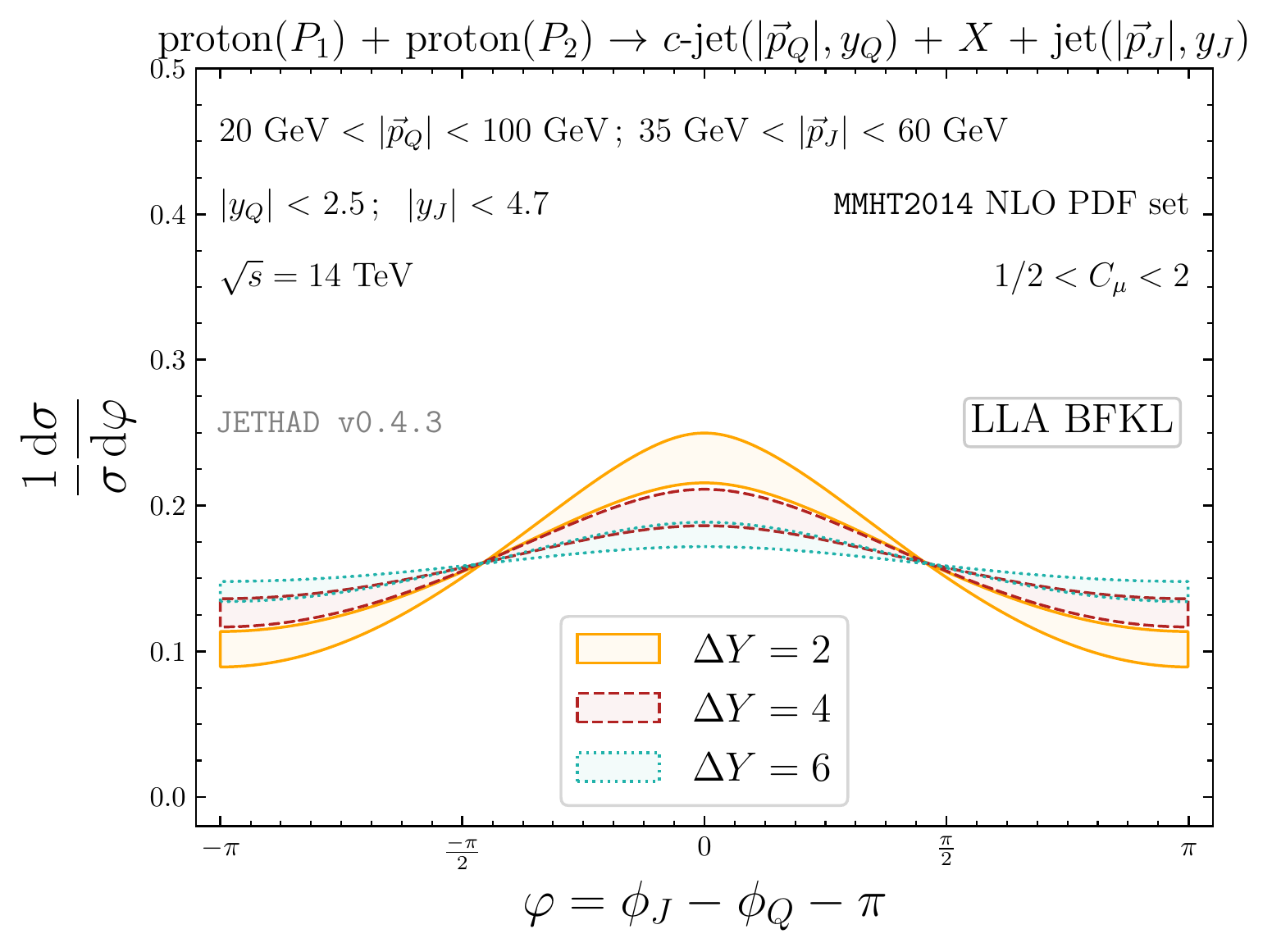}

\includegraphics[scale=0.53,clip]{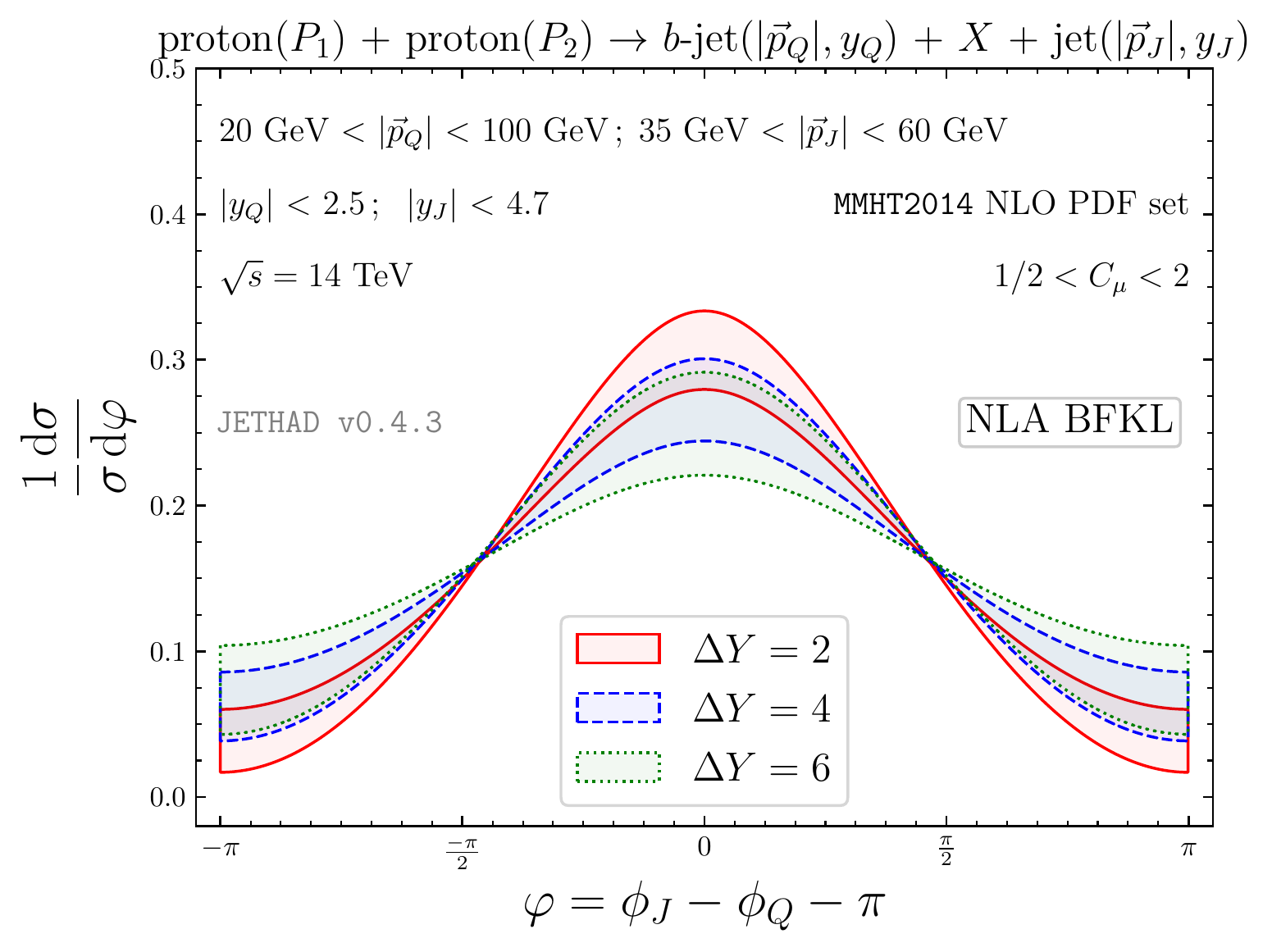}
\includegraphics[scale=0.53,clip]{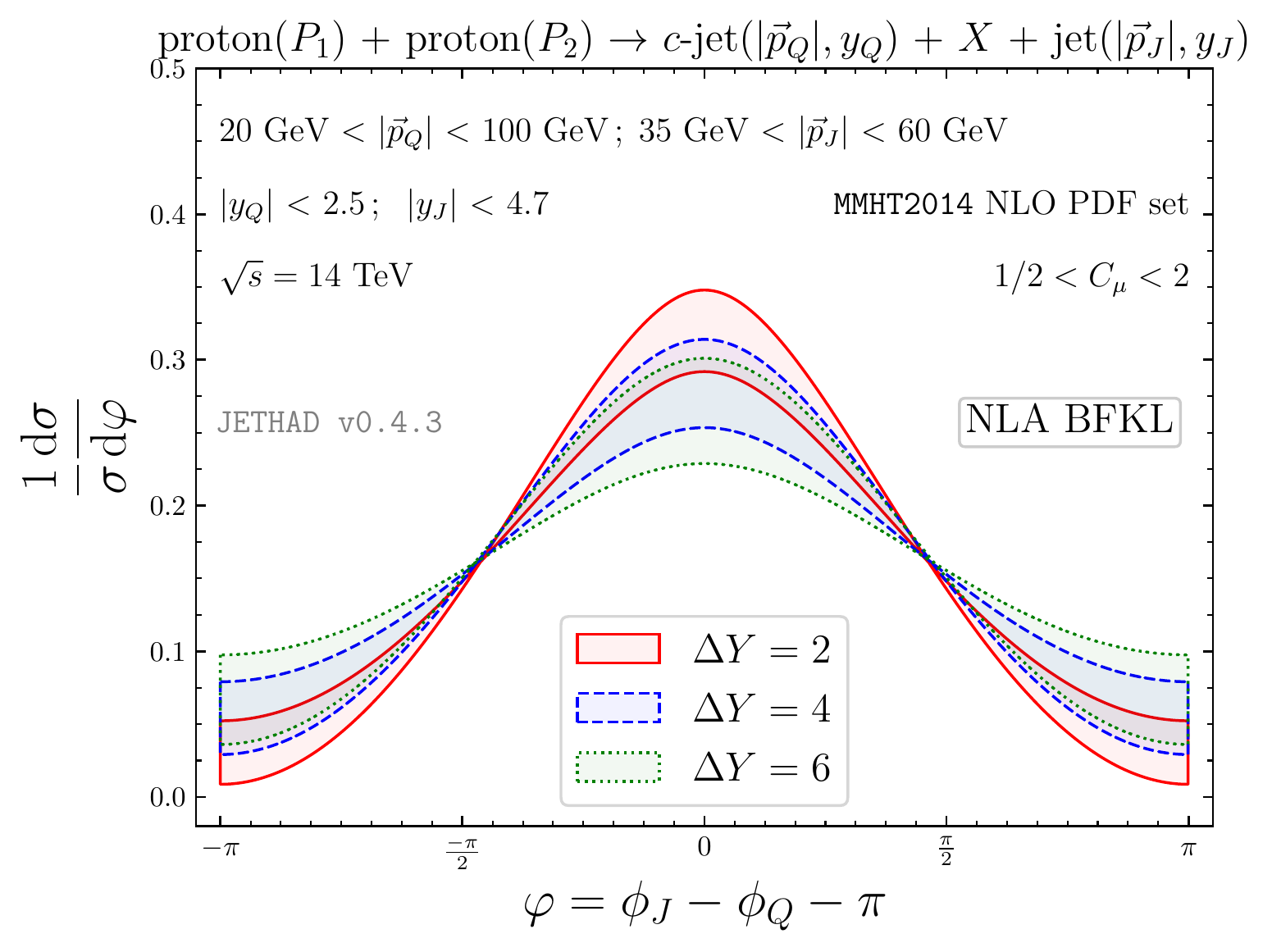}
\caption{LLA (upper) and NLA (lower) azimuthal distribution in the $b\text{-jet}$ (left) and $c\text{-jet}$ (right) channels, for three distinct values of the final-state rapidity interval, $\DY$, and for $\sqrt{s} = 14$ TeV. Shaded bands provide with the combined uncertainty coming from scale variation and numerical integration(s).}
\label{fig:azimuthal}
\end{figure}

\begin{figure}[p]
\centering
\includegraphics[scale=0.52,clip]{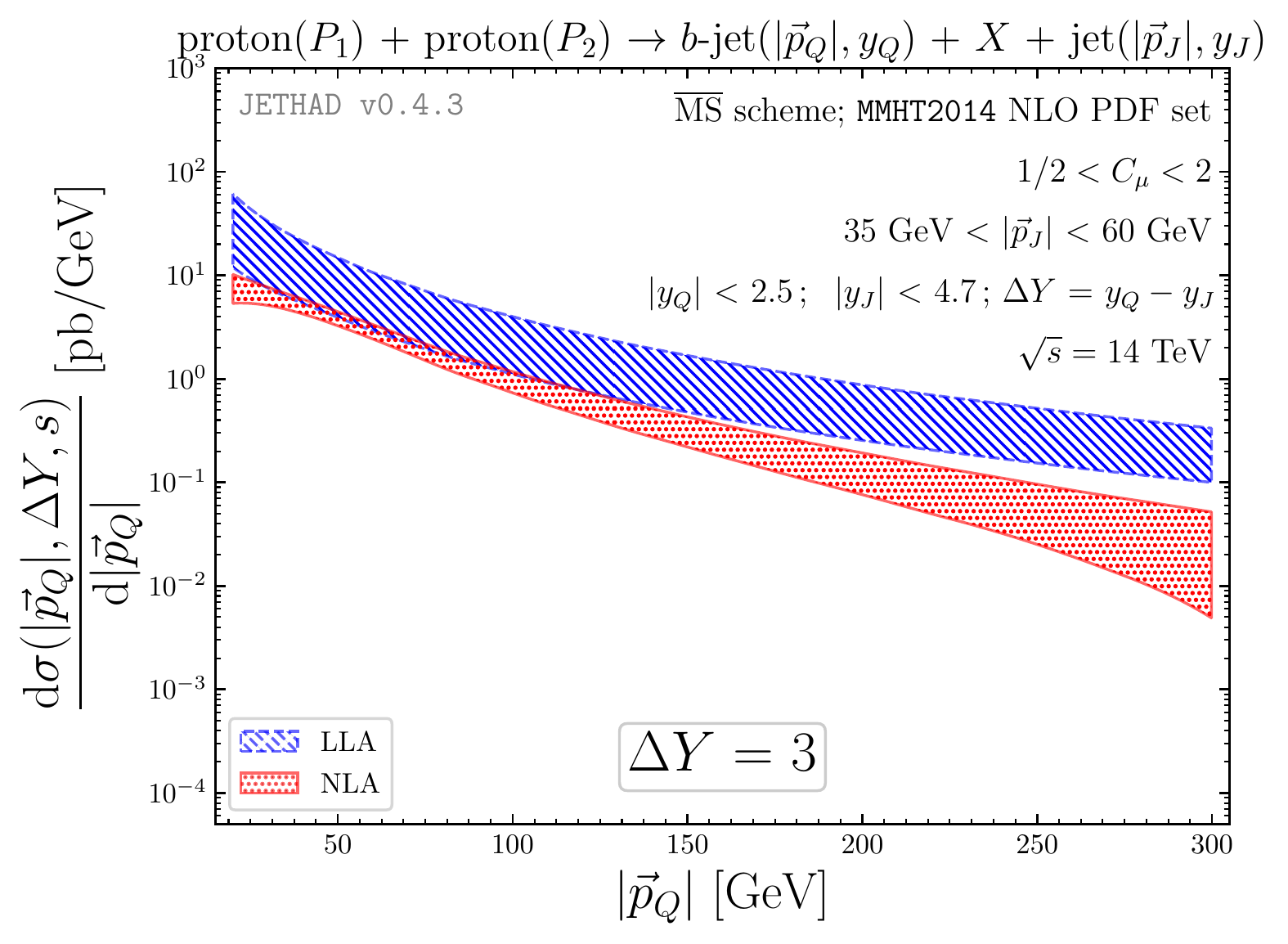}
\includegraphics[scale=0.52,clip]{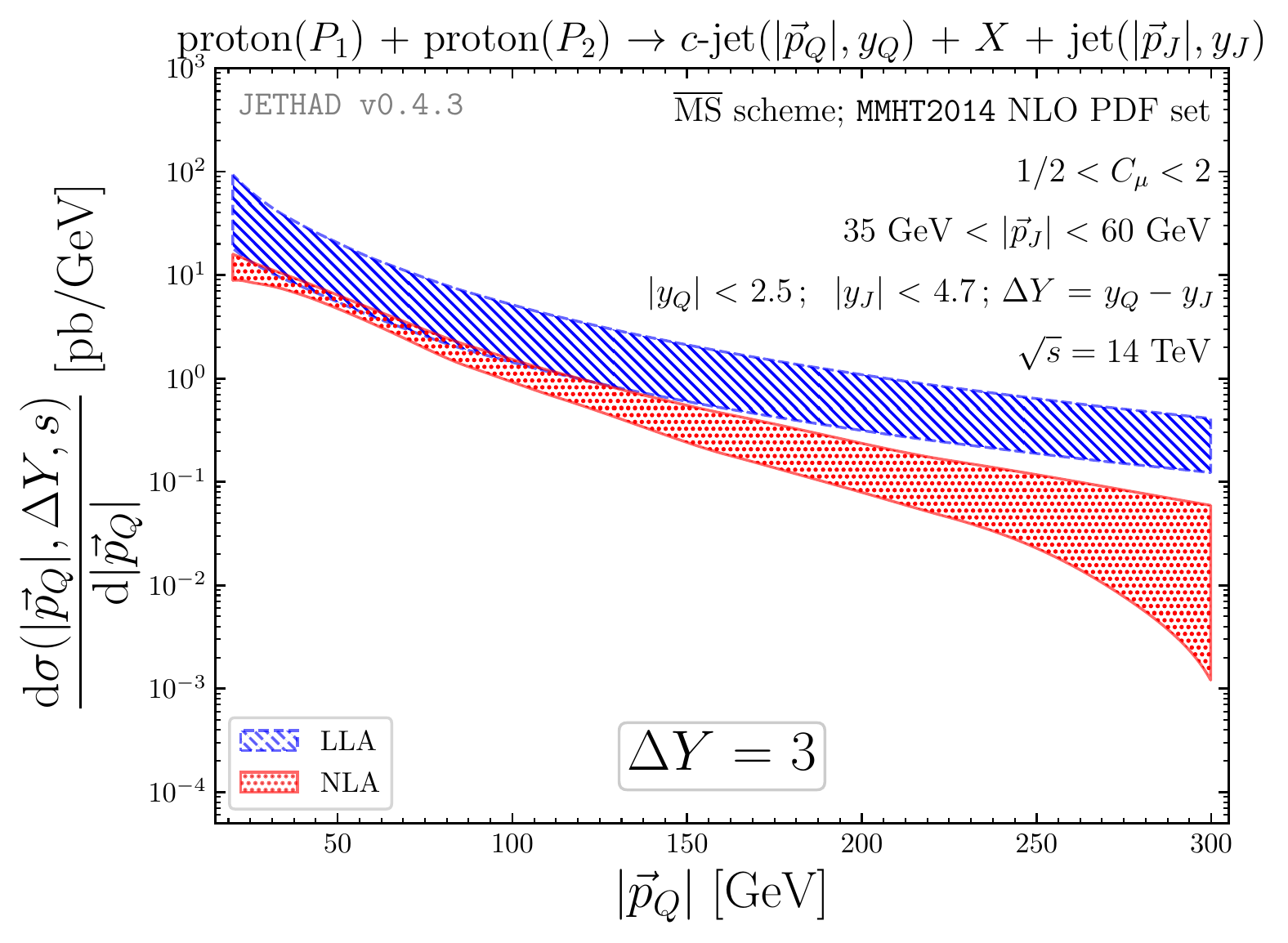}

\includegraphics[scale=0.52,clip]{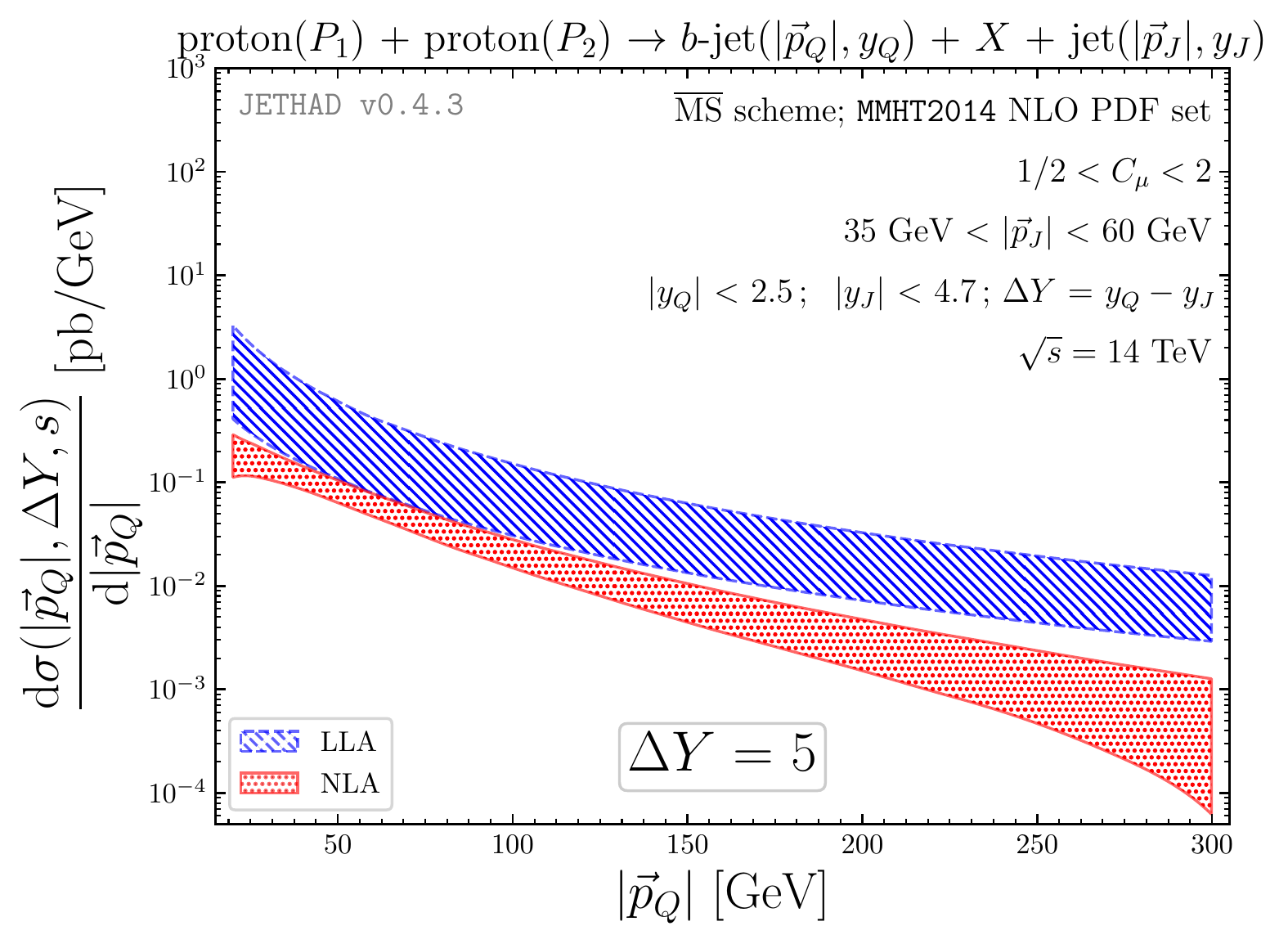}
\includegraphics[scale=0.52,clip]{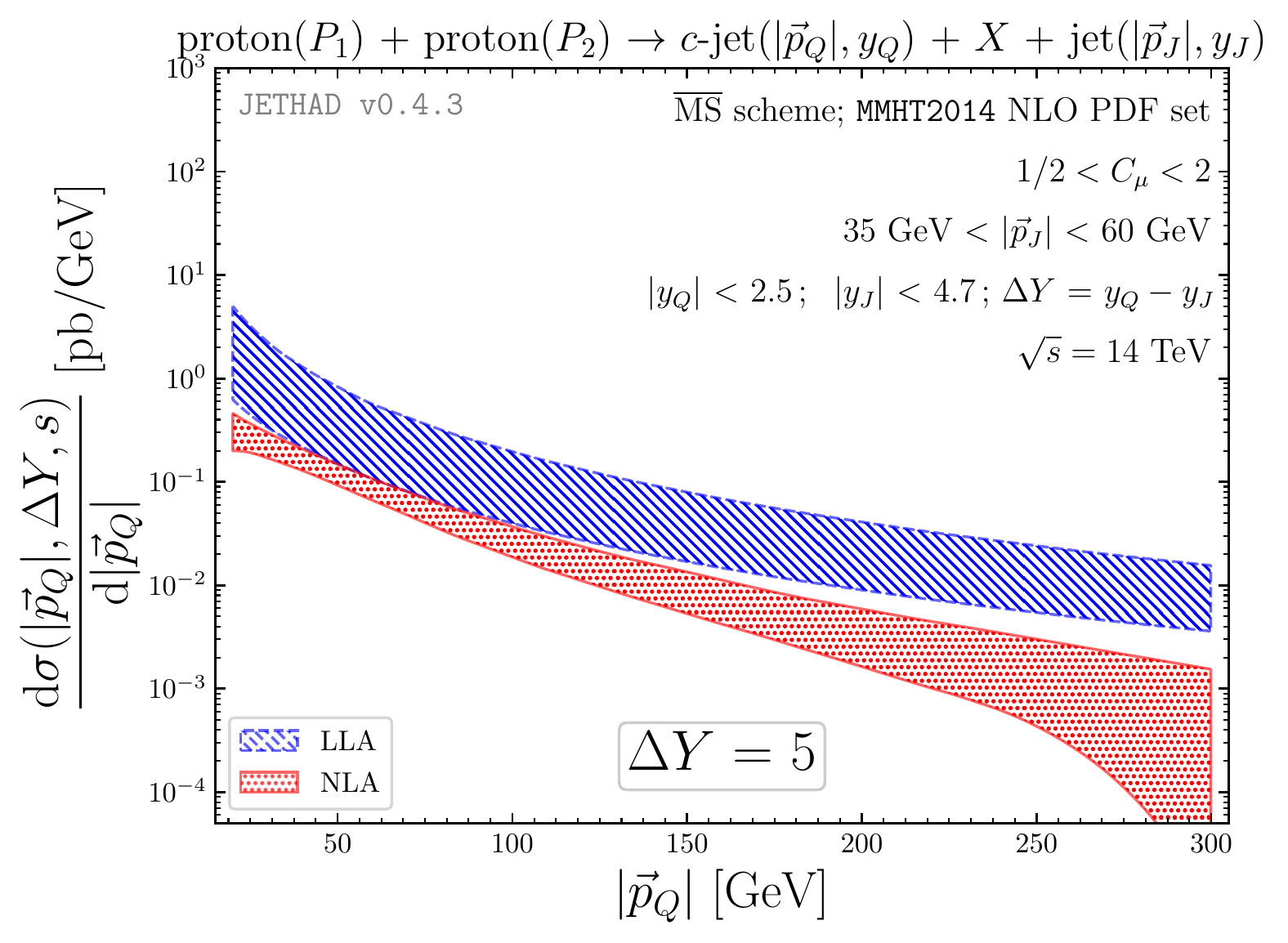}
\caption{Transverse-momentum distribution of the $b$-jet (left) and of the $c$-jet (right) at $\DY = 3$ (upper) and $\DY = 5$ (lower), for 35 GeV $< |\vec p_J| <$ 60 GeV and $\sqrt{s} = 14$ TeV. Shaded bands provide with the combined uncertainty coming from scale variation and numerical integration(s).}
\label{fig:pT}
\end{figure}

Predictions for the $\DY$-distribution, $C_0$, are presented in Fig.\tref{fig:C0}. Here, the growth with energy of partonic cross sections, predicted by BFKL, combines with the quenching effect coming from the convolution with collinear PDFs, thus resulting in a downtrend with $\DY$ of distributions. A clear manifestation of a reached stability of $C_0$ under NLA corrections is given by the size of uncertainty bands, which is sensibly lower when higher-order contributions are considered. In particular, NLA corrections to the BFKL kernel become progressively negative when $\DY$ increases, this explaining why pure LLA predictions are always larger. As an overall outcome, statistics for $\DY$-distributions is very promising, the $c\text{-jet}$ channel being larger, but of the same order of the $b\text{-jet}$ one.

Results for several azimuthal-correlation moments, $R_{nm}$, are shown in Fig.\tref{fig:Rnm_b} ($b$-jet channel) and Fig.\tref{fig:Rnm_c} ($c$-jet channel). Here, the usual onset of the high-energy dynamics clearly emerges. All azimuthal correlations fall off when $\DY$ increases, thus validating the assumption that the weight of the inclusive system of gluons emitted in the final state leads to a decorrelation of the two jets. As a result, statistics moves away from the (almost) back-to-back configuration as the available phase space provided by large rapidity intervals grows. More precisely, since the heavy-flavored jet stems from a vertex where a ${q\bar{q}}$-system is produced (see Fig.\tref{fig:hqp_IF}), there are no kinematic constrains that enforce a back-to-back event, even at LO. This explains why $R_{nm}$ ratios for the heavy-light dijet production are constantly smaller, regardless of the value of $\DY$, with respect to the case of pure light-jet or light-hadron emissions~(see, \emph{e.g.}, Refs.\tcite{Caporale:2012ih,Celiberto:2017ptm,Bolognino:2018oth,Celiberto:2020wpk}). The fact that, at variance with Mueller--Navelet jets, heavy-light dijet azimuthal correlations can be studied around natural values of renormalization and factorization scales gives us further evidence of the stability of the BFKL series, which, however is less marked than in the $C_0$ case. Here, the sensitivity to scale variation leads to an opposite situation, where NLA uncertainty bands are, at least for the ${R_{n0}}$ ratios, larger than LLA ones. This feature is not surprising, since azimuthal correlations are widely recognized to be among the most sensitive observables to high-energy dynamics.

The $\varphi$-shape of azimuthal distributions, for different values of $\DY$, is given in Fig.\tref{fig:azimuthal}. The peculiar behavior of these observables corroborates the assumption that we are probing a regime where the BFKL treatment is valid. All distributions present a distinct peak at $\varphi = 0$, namely where the two jets are emitted in back-to-back configurations. When $\DY$ increases, the peak height decreases, while the distribution width broadens. This reflects the fact that larger rapidity intervals bring to a more significant decorrelation of the dijet system, so that the number of back-to-back events diminish. The stronger decorrelation of the LLA series observed in the $R_{nm}$ patterns consistently translates in smaller peaks of the corresponding azimuthal distributions, with respect to the NLA case.

Finally, the $p_Q$-distribution at fixed rapidity distance, $\DY$, is depicted in Fig.\tref{fig:pT}. Two values of $\DY$ were considered, specifically $\DY = 3$ and $5$. For the sake of comparison with our previous analysis of the Higgs $p_T$-distribution in the inclusive Higgs-plus-jet hadroproduction, with the light-jet present in both the two cases and lying in the same transverse-momentum range (see Fig.~8 of Ref.\tcite{Celiberto:2020tmb}), we identify three contiguous kinematic subintervals. We refer to the first one as the low-$|\vec p_Q|$ region, say $|\vec p_Q| \lesssim 15$ GeV. Here, large transverse-momentum logarithms dominate, thus making our formalism inadequate. Therefore, this range was excluded from our plots. 
In the second region, $|\vec p_Q|$ is of the same order of $|\vec p_J|$, which lies in the interval between 35 and 60 GeV. It represents the most suitable range for our  description, where the high-energy resummation is expected to work properly. Indeed, NLA uncertainty bands are much narrower than LLA ones. 
Then, moving along the $|\vec p_Q|$ axis, we run into the large-$|\vec p_Q|$ region, where LLA and NLA series visibly decouples from each other and the sensitivity of the NLA one to scale variation progressively increases. In this subinterval, large DGLAP-type logarithms as well as threshold contaminations become more and more relevant, up to the point of harming the convergence of the high-energy series.

A striking difference between the heavy-flavored jet and the Higgs transverse-momentum distribution is the apparent absence of a peak in the first case, which conversely is clearly visible in the second case, both at LLA and NLA. 
We remark that, at variance with inclusive single emissions, where the peak cannot appear in fixed-order collinear calculations, but needs to be generated by the transverse-momentum resummation (see Ref.\tcite{Catani:2000vq,Bozzi:2005wk,Bozzi:2008bb,Catani:2010pd,Catani:2011kr,Catani:2013tia,Catani:2015vma} and references therein), in our two-particle final-state reactions the peak genuinely comes from the ``fixed-order part" of our calculation, the high-energy resummation simply having a modulation effect on the whole distribution.
Coming back to the loss of the peak, the ``mystery" is soon revealed.
Although gathering its position from our analytic expressions is far from straightforward (it depends also on the integration range of the $p_T$ of the light jet), numerical analyses have shown that the peak lies in the low-$|\vec p_Q|$ subregion, where, as already mentioned, our calculations are quite unstable. This is sufficient, however, to conclude that the heavy-flavored jet transverse-momentum distribution is actually similar in shape to the Higgs one, the latter being translated rightward on the $|\vec p_T|$ axis. This shift is mostly related to the different transverse masses at work, much larger in the Higgs-emission case. While the dijet peak lives outside the edges of the applicability region of the BFKL formalism, the Higgs-plus-jet peak is around $40 \div 50$ GeV, where a BFKL description is valid. As a final remark, statistics for the heavy-flavored jet transverse-momentum distribution is from two to three orders of magnitude larger than the Higgs one, thus making it a very favorable observable to be compared with forthcoming experimental analyses, at least at moderate values of $|\vec p_Q|$.

\section{Toward new directions}
\label{conclusions}

We proposed the study of the inclusive semi-hard production of a heavy-light dijet system in hybrid high-energy/collinear factorization. We performed a detailed analysis of different distributions, tailored on realistic LHC kinematic ranges and differential in the rapidity interval between the two emitted jets, in their azimuthal-angle distance, and in the transverse momentum of the heavy-flavored jet. 
Distinctive signals of the onset of the high-energy dynamics fairly emerged, thus providing us with corroborating evidence of the validity of our approach. 
We hunted for a stability of the BFKL series under higher-order corrections, discovering that, at variance with the inclusive light-dijet production (Mueller--Navelet channel), heavy-flavored emissions offer the possibility to perform studies around natural values of energy scales and, ultimately, to assess the feasibility of precision calculations at the hand of a (hybrid) high-energy treatment. As expected, due to the lower transverse masses at work, these stabilizing effects are more moderate with respect to the inclusive Higgs-plus-jet case\tcite{Celiberto:2020tmb}.
Future investigations will extend our work to a full NLA BFKL analysis and to a comparison with fixed-order calculations.

The studies proposed constitute a step forward in our ongoing program on heavy-flavored emissions at high energies, started from the analytic calculation of heavy-quark pair impact factors\tcite{Celiberto:2017nyx,Bolognino:2019ouc,Bolognino:2019yls} and pointing toward the analysis of quark bound states, as heavy-light mesons and quarkonia\tcite{Boussarie:2017oae,Arbuzov:2020cqg,Chapon:2020heu}. In this context, both two-particle and single-particle emissions are relevant. In particular, the second case will allow us to deepen our knowledge of the already mentioned small-$x$ UGD, a key ingredient to describe single forward as well as central emissions in the high-energy regime (Figs.\tref{fig:process_2p} and\tref{fig:process_1p}).
Moreover, it offer us the chance of comparing different approaches of hybrid factorization at small-$x$ (see, \emph{e.g.}, Refs.\tcite{vanHameren:2015uia,Deak:2018obv}), and possibly realizing joint analyses.

Another intriguing perspective is represented by the exploration of wider kinematic domains, where other mechanisms become relevant. As pointed out in our study on the $p_Q$-distribution (see Section\tref{results}), a more solid description would rely on a \emph{multi-lateral} formalism where effects coming from other resummations are consistently embodied. We mainly refer to threshold, Sudakov and low-$p_T$ effects.

The advent of the Electron-Ion Collider (EIC)\tcite{EICUGYR:2020} will open a new phase in the search for New Physics via the study of the hadronic structure. Here, heavy-flavored jet emissions act as powerful probes of a wide kinematic spectrum, in many aspects complementary with respect to the LHC one. In order to complement these analyses by evaluating the size of high-energy/small-$x$ effects, we will need to enhance our hybrid factorization. A first possibility would be encoding low-$p_T$ features in our formalism. In this respect, the improvement of the BFKL description via the inclusion of the Sudakov resummation of small-$p_T$ imbalances in almost back-to-back emissions, proposed in Refs.\tcite{Mueller:2012uf,Mueller:2015ael,Xiao:2018esv}, could serve as a useful guidance. Another challenging option would consist in starting from a genuine TMD formalism and plugging small-$x$ effects on top of it. More phenomenological ways would rely on using a hybrid factorization with small-$x$ improved TMD densities\tcite{Bacchetta:2020vty} or with CGC/JIMWLK distributions (see, \emph{e.g.}, Refs.\tcite{Dominguez:2011wm,Kotko:2015ura,Marquet:2016cgx,Marquet:2017xwy,Altinoluk:2018byz}). 
Moreover, at EIC regimes, the distortion of the isotropy of soft-gluon radiation should be encoded in the description of (heavy) jet emission, this leading to a more sophisticated treatment of the jet-selection function (see, \emph{e.g.}, Refs.\tcite{Gutierrez-Reyes:2019msa,Arratia:2020ssx,Makris:2021drz}).
Finally, final-state sensitivity to the jet substructure in terms of heavy hadrons within jets\tcite{Procura:2009vm,Baumgart:2014upa,Bain:2017wvk} should be, if possible, gauged.

All these engaging opportunities constitute the bulk of our \emph{second direction}, namely the use of the high-energy resummation as an additional tool to examine observables sensitive to heavy-flavored emissions.
We propose these studies with the aim of inspiring synergies with other Communities, and pursuing the goal of widening common horizons in the exploration of heavy-flavor physics.

\section*{Acknowledgements}

We thank V. Bertone, G. Bozzi, M.G. Echevarr\'ia, and P. Taels for fruitful conversations.
A.D.B., M.F. and A.P. acknowledge support from the INFN/QFT@COLLIDERS project.
F.G.C. acknowledges support from the INFN/NINPHA project and thanks Universit\`a degli Studi di Pavia for the warm hospitality.
The work of D.I. was carried out within the framework of the state contract of the Sobolev Institute of  Mathematics (Project No. 0314-2019-0021).

\bibliographystyle{apsrev}
\bibliography{bibliography}

\end{document}